\documentclass[reprint, aps, prl]{revtex4-2}

\usepackage{graphicx}
\usepackage{amsmath,amssymb,bm}
\usepackage{xcolor}
\usepackage{braket}
\usepackage{physics}
\usepackage{dcolumn}

\usepackage{textcomp}
\usepackage[normalem]{ulem}

\usepackage[colorlinks]{hyperref}

\begin{document}

\title{Optically tunable nonlinear mechanical damping in an optomechanical resonator}

\author{Hideki Arahari}
\email[]{hideki.arahari@ntt.com}
\thanks{These authors contributed equally to this work}
\author{Motoki Asano}
\email[]{motoki.asano@ntt.com}
\thanks{These authors contributed equally to this work}
\author{Hiroshi Yamaguchi}
\author{Hajime Okamoto}

\affiliation{
 Basic Research Laboratories, NTT, Inc., 
 3-1 Morinosato Wakamiya, Atsugi-shi, Kanagawa 243-0198 Japan
}

\date{\today}

\begin{abstract}
We theoretically propose and experimentally demonstrate optically tunable nonlinear mechanical damping in a cavity optomechanical system utilizing a partly resolved sideband regime. Optomechanical coupling provides a delayed nonlinear backaction to the mechanical modes, resulting in nonlinear mechanical damping. This optically induced nonlinear damping is observed in the frequency and time domains, and we show using both theory and experiment that it can be tuned
via laser detuning. Specifically, we observe positive nonlinear damping in the blue-detuned regime and negative nonlinear damping in the red-detuned regime. We also observe optically mediated cross-nonlinear damping between two mechanical modes: the amplitude of one mode modulates the damping of the other. The presented results show a tunable scheme of nonlinear mechanical damping that will be applicable to various non-trivial systems, governed by nonlinear, nonequilibrium, and non-Hermitian phenomena.
\end{abstract}

\maketitle

\textit{Introduction -} Nonlinear damping, where the rate of energy dissipation depends on the system dynamics, is a ubiquitous phenomenon that appears across a wide range of oscillatory systems including ship dynamics \cite{taylan2000effect, yang2013trajectory}, electrical circuits \cite{cartwright1960balthazar}, and biological rhythms \cite{fitzhugh1961impulses}. Beyond such macroscopic examples, nonlinear damping also plays a significant role in well-engineered mesoscale platforms such as mechanical resonators \cite{buks2006mass,Lifshitz2009,zaitsev2012nonlinear,imboden2013observation,Eichler2011, guttinger2017energy,Singh2016,yanai2017mechanical,dong2018strong,catalini2020soft,Catalini2021,kecskekler2021tuning,miller2021amplitude}, superconducting circuits \cite{gely2023apparent}, and nanomagnets \cite{barsukov2019giant}. In these systems, amplitude-dependent dissipation governs key dynamical behaviors including self-stabilization of oscillation amplitudes \cite{miller2021amplitude}, intermodal coupling \cite{catalini2020soft}, and dissipative reservoir engineering for tailoring quantum states \cite{gely2023apparent}.

At a phenomenological level, nonlinear damping is often captured by a cubic friction term, as exemplified by the van der Pol equation \cite{van1926lxxxviii}. Over the past decade, substantial progress has been made toward elucidating microscopic mechanisms underlying nonlinear damping in mesoscale devices and this has clarified how effective nonlinear damping emerges from device geometry \citep{catalini2020soft,Catalini2021}, material properties \citep{atalaya2016nonlinear}, and mode coupling \citep{Mahboob2015, guttinger2017energy, kecskekler2021tuning, prasad2023tunable}. However, in most existing platforms, such nonlinear damping is intrinsically fixed once the system is fabricated rendering it difficult to tune in situ. This lack of controllability has hindered systematic exploration of nonlinear dissipation as a universal nonequilibrium resource, which limits access to phenomena such as bifurcation dynamics \cite{dykman2007critical}, parity-time symmetric dynamics \cite{karthiga2016twofold, ramezani2010unidirectional}, and collective nonequilibrium dynamics stabilized by irreversible phase-space flows \cite{seifert2012stochastic}.

In contrast, Refs. \cite{dong2018strong, chan2026multibranched} report a common method for controlling nonlinear damping in mechanical two-mode systems. This approach achieves controlled negative nonlinear damping by utilizing a distinct ``faster-decaying mode" as the damping source. Inspired by their work, we here propose an alternative technique to control nonlinear damping simply through a small detuning of the pump excitation in dispersively coupled cavity-resonator systems. We demonstrate optically tunable nonlinear mechanical damping arising from delayed optical backaction in a cavity optomechanical system. In a partly resolved sideband regime where the cavity response time is comparable to the mechanical oscillation period, nonlinear damping is selectively tuned by laser detuning. We identify the physical origin of this effect as the phase shift introduced by the finite cavity response, which converts delayed nonlinear backaction into amplitude-dependent dissipation. We further extend this framework to multiple mechanical modes and experimentally reveal optically mediated cross-nonlinear damping as a purely dissipative interaction. These results establish cavity optomechanics as a versatile platform for engineering nonlinear dissipation and nonequilibrium mechanical dynamics. More broadly, because delayed dispersive interactions are ubiquitous in cavity-based systems, the underlying concept should be applicable to other physical systems~\cite{teufel2008dynamical, potts2021dynamical, rodrigues2021cooling}.

\begin{figure*}[t]
\centering
\includegraphics[width=\linewidth]{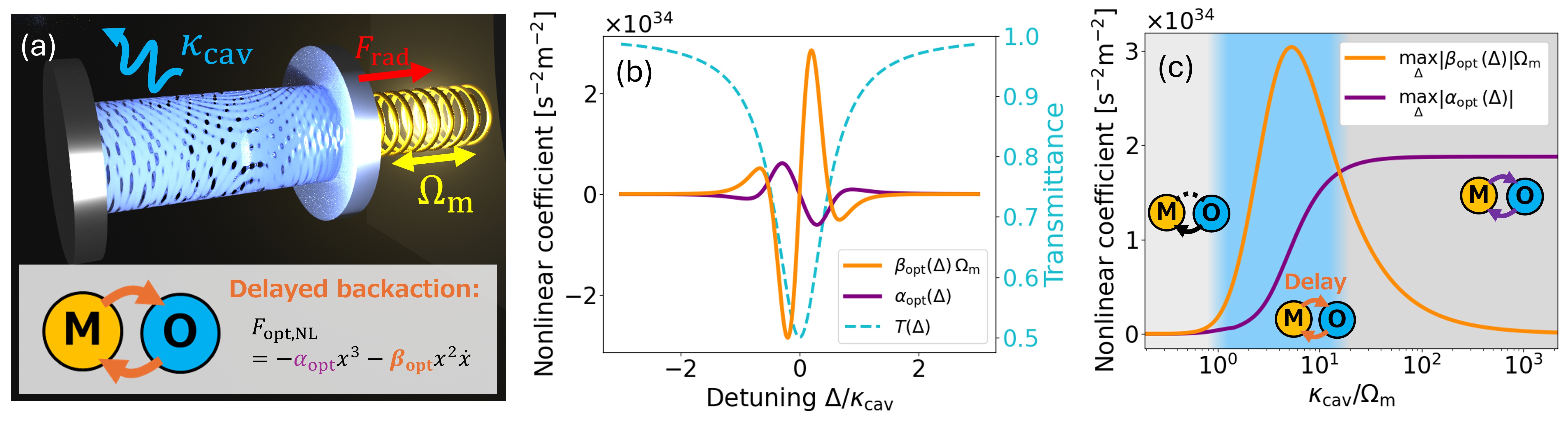}
\caption{
(a) Concept of nonlinear mechanical damping induced by delayed backaction in cavity optomechanics, where cavity decay rate $\kappa_{\mathrm{cav}}$ becomes comparable to mechanical frequency $\Omega_{\mathrm{m}}$. 
(b) Theoretical detuning dependence of the mechanical nonlinearities near cavity resonance. The orange curve shows nonlinear damping strength $\beta_{\mathrm{opt}}(\Delta)\Omega_{\mathrm{m}}$ and the purple curve represents Duffing nonlinearity $\alpha_{\mathrm{opt}}(\Delta)$. The light-blue dashed line indicates the optical cavity transmittance. 
(c) Theoretically calculated maximum strength of the nonlinearities optimized over detuning $\Delta$ as a function of $\kappa_{\mathrm{cav}}/\Omega_\mathrm{m}$. The orange curve represents maximum nonlinear damping strength $\max_{\Delta}\abs{\beta_{\mathrm{opt}}(\Delta)}\Omega_{\mathrm{m}}$ and the purple curve shows maximum Duffing nonlinearity $\max_{\Delta}\abs{\alpha_{\mathrm{opt}}(\Delta)}$. Schematic illustrations highlight the distinct optomechanical backaction mechanisms in different parameter regimes.
All calculations use parameters corresponding to the employed experimental system (a microbottle optomechanical resonator, as discussed later in Fig.~\ref{fig:2}) assuming an input optical power of 50 mW and measured optomechanical coupling $g/2\pi=29.6$ Hz, determined using the method of Ref.~\cite{gorodetksy2010determination}. 
}
\label{fig:1}
\end{figure*}

\textit{Theoretical model - }Optomechanical coupling induces mechanical nonlinearity through a nonlinear backaction force, whereby a strong optical field reflects forces containing higher-order terms in the mechanical displacement \cite{Aspelmeyer2014,Rodrigues2021,Asano2025}. Such a nonlinear backaction force can acquire a non-conservative component - namely, nonlinear damping - when the cavity conditions are tuned to introduce a finite delay, as shown in Fig.~\ref{fig:1}(a). The magnitude of this delay is governed by resolved sideband parameter $\kappa_{\mathrm{cav}}/\Omega_\mathrm{m}$, where $\kappa_\mathrm{cav}$ and $\Omega_\mathrm{m}$ denote the optical cavity dissipation rate and mechanical resonance frequency, respectively. 

To illustrate this behavior, we express the mechanical displacement as $x(t)\simeq x_0\cos(\Omega_{\mathrm{m}}t)$ and represent the delayed backaction by a phase lag
$\phi=\Omega_{\mathrm{m}}\tau$. Denoting the nonlinear component of the radiation-pressure force by $F_{\mathrm{rad,NL}}$, its fundamental-frequency component gives
\begin{equation}
\begin{aligned}
F_{\mathrm{rad,NL}}(t)
&\propto x_0^3\cos^3(\Omega_{\mathrm{m}}t-\phi) \\
&\simeq
\frac{3x_0^2}{4}
\left[
\cos\phi~x(t)
-\frac{\sin\phi}{\Omega_{\mathrm{m}}}\dot{x}(t)
\right].
\end{aligned}
\label{eq:delayed_nonlinear_force}
\end{equation}
The displacement- and velocity-terms give rise to nonlinear stiffness and nonlinear damping, respectively, with their common $x_0^2$ prefactor reflecting the cubic nonlinearity. In the adiabatic regime, $\kappa_{\mathrm{cav}}\gg\Omega_{\mathrm{m}}$, the phase lag is negligible and the nonlinear stiffness term dominates. By contrast, in the partially resolved-sideband regime, $\kappa_{\mathrm{cav}}\sim\Omega_{\mathrm{m}}$, the finite phase lag produces an appreciable velocity component, whose magnitude and sign depend on the cavity conditions.

We theoretically quantify nonlinear backaction effects on the $j$th mechanical mode, $b_j$ $(j=1,2,\cdots, N)$, induced by an optical cavity field using a complex-amplitude description under the rotating-wave approximation. The equations of motion for both optical and mechanical amplitudes, $a(t)$ and $b_j(t)$, are, respectively, given by
\begin{eqnarray}
    &&\dot{a}(t) = \left[i \Delta - \frac{\kappa_{\mathrm{cav}}}{2}\right]a(t) \nonumber\\
    &&\qquad+ i\sum_{j=1}^N g_j a(t)\left(b_j(t)+b_j^\ast(t)\right) + \sqrt{\kappa_{\mathrm{in}}}a_{\mathrm{in}},\label{eq:Opt}\\
    &&\dot{b}_j(t) = -i\Omega_{j}b_j(t) -\frac{\Gamma_j}{2}b_j(t) + ig_j |a(t)|^2 . \label{eq:Mech}
\end{eqnarray}
Here, $\Delta$ is the laser detuning from the cavity resonance, $\kappa_{\mathrm{cav}}$ is the total cavity decay rate, and $\kappa_{\mathrm{in}}$ is the coupling rate to input optical field $a_\mathrm{in}$. For the $j$th mechanical mode, $\Omega_{j}$, $\Gamma_j$, and $g_j$ denote the resonance frequency, damping rate, and optomechanical coupling rate, respectively. A general formulation of nonlinear optomechanical backaction is obtained by expressing the formal solution of Eq.~(\ref{eq:Opt}) as a time-domain integral and expanding it to include higher-order perturbative contributions. This procedure incorporates the history of the mechanical motion into the radiation-pressure response through a memory kernel, yielding an explicit third-order expression for the optically induced nonlinearities [see Supplemental Material for details]. For simplicity, here we show the equation of motion for a single mechanical mode $N=1$ given by 
\begin{eqnarray}
    \dot{b} 
    &=& -i\left(\Omega_{\mathrm{eff}}
        - i\frac{\Gamma_{\mathrm{eff}}}{2}\right)b \nonumber\\
    && - i\left(\frac{3\alpha_{\mathrm{opt}}}{2\Omega_{\mathrm{m}}}
        - i\frac{\beta_{\mathrm{opt}}}{2}\right)x_{\mathrm{zpf}}^2|b|^{2} b,
\end{eqnarray}
where $\Omega_{\mathrm{eff}}$ and $\Gamma_{\mathrm{eff}}$, respectively, denote the effective mechanical resonance frequency and damping modified by linear optomechanical backaction, and $x_{\mathrm{zpf}}$ is the zero-point fluctuation amplitude of the mechanical oscillator.
The equation includes both real and imaginary contributions of the nonlinear backaction force with the coefficients of $\alpha_\mathrm{opt}$ and $\beta_\mathrm{opt}$, respectively, with the imaginary part representing optically induced nonlinear damping. Each coefficient explicitly depends on the detuning, as follows:
\begin{eqnarray}
    \alpha_{\mathrm{opt}}(\Delta)
    &=& \frac{2\Omega_{\mathrm{m}}}{3}
        \frac{n_{\mathrm{cav}}(\Delta)g^{4}}{x_{\mathrm{zpf}}^{2}}
        \Re[C(\Delta)],\\
    \beta_{\mathrm{opt}}(\Delta)
    &=& -2\frac{n_{\mathrm{cav}}(\Delta)g^{4}}{x_{\mathrm{zpf}}^{2}}
        \Im[C(\Delta)],\label{eq:5}\\
    C(\Delta) &=& i\left[S_{+}(\Delta)-S_{-}^{*}(\Delta)\right], \\
    S_{\pm}(\Delta) &=& \chi_{\pm}\chi_{0}\chi_{\mp} + \left(\chi_{\pm}\right)^2 \chi_{0} + \left(\chi_{\pm}\right)^2\chi_{\pm2} \nonumber\\
    &&+ \chi_{\mp}\chi^{*}_{\mp2} \chi^{*}_{\mp} 
    + \chi_{\pm} \chi^{*}_{0} \left(\chi_{\pm} 
    + \chi_{\mp} \right)^*.
\end{eqnarray}
Here, $n_{\mathrm{cav}}(\Delta)=\kappa_{\mathrm{in}} |a_{\mathrm{in}}|^2 |\chi_0(\Delta)|^2$ denotes the bare intracavity photon number, and $\chi_{\pm n} = \left(\kappa_{\mathrm{cav}}/2 - i(\Delta\pm n \Omega_{\mathrm{m}}) \right)^{-1}$ describes the first-order optomechanical susceptibility, where $n$ is an integer.

\begin{figure*}[t]
\centering
\includegraphics[width=\linewidth]{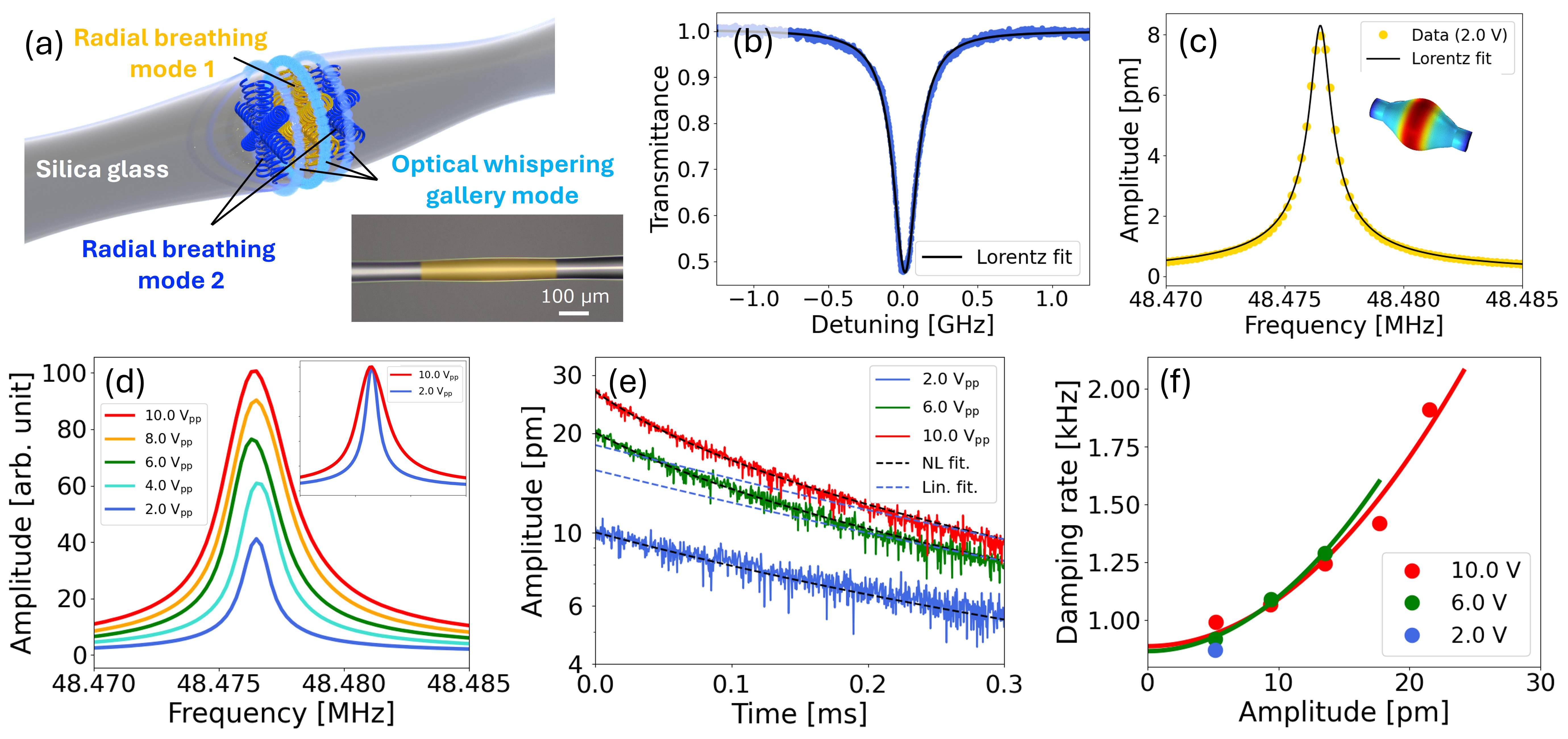}
\caption{
(a) Schematic of microbottle optomechanical resonator and its optical micrograph, in which the bottle region is highlighted in light yellow. Two representative radial breathing modes are illustrated, anticipating the multimode discussion in Fig.~\ref{fig:4}.
(b) Optical transmission spectrum (at $\lambda = 1.55~\mathrm{\mu m}$) with a Lorentzian fit yielding $\kappa_{\mathrm{cav}}/2\pi = 194$~MHz ($Q_{\mathrm{cav}} = 1.0 \times 10^{6}$).
(c) Driven mechanical frequency response of a representative radial breathing mode (RBM) under weak excitation with a Lorentzian fit yielding $\Gamma_{\mathrm{m}}/2\pi = 0.83$ kHz at $\Omega_{\mathrm{m}}/2\pi = 48.476$ MHz.
(d) Driven mechanical response at different excitation strengths parameterized by the voltage applied to the EOM, $V_{\mathrm{EOM}}$. Inset: Corresponding normalized spectra at 10.0 V and 2.0 V. 
(e) Ring-down traces for different drive strengths. Black dashed curves show nonlinear-damping fits while blue dashed curves show linear-damping fits. Fit yields $\Gamma_{\mathrm{0}}/2\pi=0.91\ \mathrm{kHz}$ and $\beta_{\mathrm{opt}}/2\pi=1.1\times10^{25}\ \mathrm{Hz}/\mathrm{m}^2$.
(f) Damping rate versus oscillation amplitude, exhibiting quadratic scaling consistent with cubic nonlinear damping.  The absolute amplitude is calibrated from the thermal motion \citep{karabalin2009piezoelectric} of the RBM using effective mass $m_{\mathrm{eff}} \simeq  7\times 10^{-9}$ kg (see Supplemental Material for details).
}
\label{fig:2}
\end{figure*}

We first examine the detuning dependence of $\alpha_\mathrm{opt}(\Delta)$ and $\beta_\mathrm{opt}(\Delta)$ for $\kappa_\mathrm{cav}/\Omega_{\mathrm{m}}= 4.0$ [see Fig.~\ref{fig:1}(b)]. At this parameter, the nonlinear damping clearly dominates the Duffing nonlinearity across a wide detuning range. In terms of their detuning dependence, similar to the first-order backaction effect, {\it e.g.}, optical spring effect, both coefficients attain their maxima near the slope of the cavity resonance. The nonlinear coefficients are determined by the third derivative of the cavity response with respect to detuning, in contrast to the first-derivative dependence of linear optomechanical backaction. Notably, enhanced positive (negative) nonlinear damping emerges on the blue (red)-detuned side. This behavior is opposite to that of linear optomechanical damping (see Fig.~S4 of the Supplemental Material) and  originates from the higher-order phase accumulation associated with delayed nonlinear backaction, resulting in a nonlinear dissipative force with reversed sign.
We then evaluate $\max_{\Delta}|\alpha_\mathrm{opt}(\Delta)|$ and $\max_{\Delta}|\beta_\mathrm{opt}(\Delta)|\Omega_{\mathrm{m}}$ as functions of $\kappa_\mathrm{cav}/\Omega_{\mathrm{m}}$ to identify the conditions for observing optically induced nonlinear damping [Fig.~\ref{fig:1}(c)]. In adiabatic regime $\kappa_\mathrm{cav}/\Omega_{\mathrm{m}} \gg1$, the mechanical nonlinearity is
predominantly Duffing-like, as the delayed backaction becomes negligible. In contrast, when $\kappa_\mathrm{cav}$ becomes comparable to
$\Omega_{\mathrm{m}}$, the mechanical nonlinearity is dominated by nonlinear damping, reflecting the emergence of nonlinear damping induced by the finite backaction delay through the optical cavity. For non-adiabatic limit $\kappa_\mathrm{cav}/\Omega_{\mathrm{m}}\ll1$, the backaction effect is suppressed because the optical cavity cannot respond to variations in the mechanical modes. 

\textit{Nonlinear damping in a single mechanical mode - }The optically induced nonlinear damping reported here arises generically in cavity optomechanical resonators operated in the partly resolved sideband regime, $\kappa_\mathrm{cav} \sim \Omega_{\mathrm{m}}$. As a concrete experimental actualization, we employ a microbottle optomechanical resonator fabricated on a silica fiber via the heat-and-pull technique \cite{sumetsky2004whispering,pollinger2009ultrahigh,Asano2016,macdonald2016optomechanics}. The resonator supports a dense spectrum of optical whispering-gallery
modes (WGMs) with various Q values, enabling the selective excitation of a specific optical resonance with a desired value of $\kappa_{\mathrm{cav}}$ and thereby providing a versatile platform for accessing $\kappa_\mathrm{cav} \sim \Omega_{\mathrm{m}}$. The maximum diameter, neck diameter, and the length between the two necks of the microbottle were 80 $\mathrm{\mu m}$, 70 $\mathrm{\mu m}$, and 450 $\mathrm{\mu m}$, respectively, as shown in Fig.~\ref{fig:2}(a). To couple laser light evanescently into an optical WGM, a tapered optical fiber with the diameter of approximately 1.5~$\mathrm{\mu m}$ was brought into contact with the resonator. By scanning the laser frequency and measuring the transmission spectra of the optical WGMs, we obtained high-Q resonances with optical quality factors $Q_\mathrm{opt}\sim 10^6$ [see Fig.~\ref{fig:2}(b) for a representative spectrum]. 

The nonlinear mechanical damping was investigated in the radial breathing modes (RBMs) of the microbottle resonator. We emphasize that the RBMs exhibit negligible geometric nonlinearity because their displacement amplitudes (in the order of picometers) are much smaller than the device dimensions (sub-mm diameter). The geometric Duffing coefficient, $\alpha_{\mathrm{geo}}$, is estimated in the Supplemental Material and is more than eight orders of magnitude smaller than the optically induced Duffing coefficient, $\alpha_{\mathrm{opt}}$.

To observe mechanical resonance characteristics, two lasers at wavelengths of 1550~nm (pump) and 1520~nm (probe) were simultaneously injected into different optical modes. The pump laser, with a power of approximately 50 mW, serves two roles: it induces optomechanical nonlinearity and drives the mechanical modes via intensity modulation using an electro-optic modulator (EOM). The probe laser was tuned to the slope of an optical resonance to measure independently the mechanical vibration using a lock-in amplifier. All experiments were performed at ambient pressure and room temperature. The configuration details are given in the Supplemental Material. Figure~\ref{fig:2}(c) shows the measured frequency response of the RBMs, with the resonance frequency of approximately 48.5 MHz and the mechanical damping rate of $\Gamma_{\mathrm{m}}/2\pi = 0.83$~kHz, in
agreement with previous reports \cite{Asano2016,Asano2022,Asano2025}. Under these conditions, the resolved sideband parameter is $\kappa_{\mathrm{cav}}/\Omega_{\mathrm{m}} \simeq 4.0$, where nonlinear damping dominates, which is consistent with Fig.~\ref{fig:1}(c).

To verify the emergence of nonlinear mechanical damping, we observed both the frequency and temporal responses with respect to the mechanical drive strength, {\it i.e.}, the applied radio-frequency (RF) voltage to the EOM, $V_\mathrm{EOM}$, under blue-detuned conditions. Figure~\ref{fig:2}(d) shows the driven mechanical response measured using lock-in detection at different $V_\mathrm{EOM}$ values. As the excitation strength increases, the resonance linewidth increases markedly, with the oscillation amplitude growing, which is a clear sign of nonlinear mechanical damping. In addition to the frequency responses, ring-down measurements were performed to measure directly the mechanical decay rate as a function of oscillation amplitude. The RF signal applied to the EOM was abruptly switched off to perform the ring-down measurement and we directly measured the photodetector output without a lock-in to avoid additional signal fluctuations introduced by the amplifier process. The resulting signal exhibited exponential-like tails corresponding to the mechanical decay as shown in Fig.~\ref{fig:2}(e). At weak excitation (blue trace) the decay is purely exponential with a constant slope; from this slope the linear mechanical dissipation rate was estimated to be $\Gamma_{\mathrm{0}}/2\pi =0.91~\mathrm{kHz}$. Under stronger excitation (green and red traces) the decay becomes faster at large amplitudes and slower at small amplitudes, which is clear evidence of nonlinear damping. Fitting the ring-down dynamics to the nonlinear-damping model yields the nonlinear damping strength $\beta_{\mathrm{opt}}\Omega_{\mathrm{m}} = 2.1\times10^{34}~\mathrm{s^{-2}m^{-2}}$. This value shows quantitative agreement with the theoretical estimate for $\beta_{\mathrm{opt}}\Omega_{\mathrm{m}}$ shown in Fig.~\ref{fig:1}(b). From the profile of Fig.~\ref{fig:2}(e), we extracted the dissipation rate as a function of the amplitude [see Fig.~\ref{fig:2}(f)]. We obtain the maximum dissipation rate of $\Gamma_{\mathrm{max}}/2\pi = 1.9\ \mathrm{kHz}$, which is more than twice the linear damping rate $\Gamma_{\mathrm{0}}$. The damping rate scales quadratically with the amplitude, which confirms a cubic nonlinear damping force.

\begin{figure*}[tb]
\centering
\includegraphics[width=\linewidth]{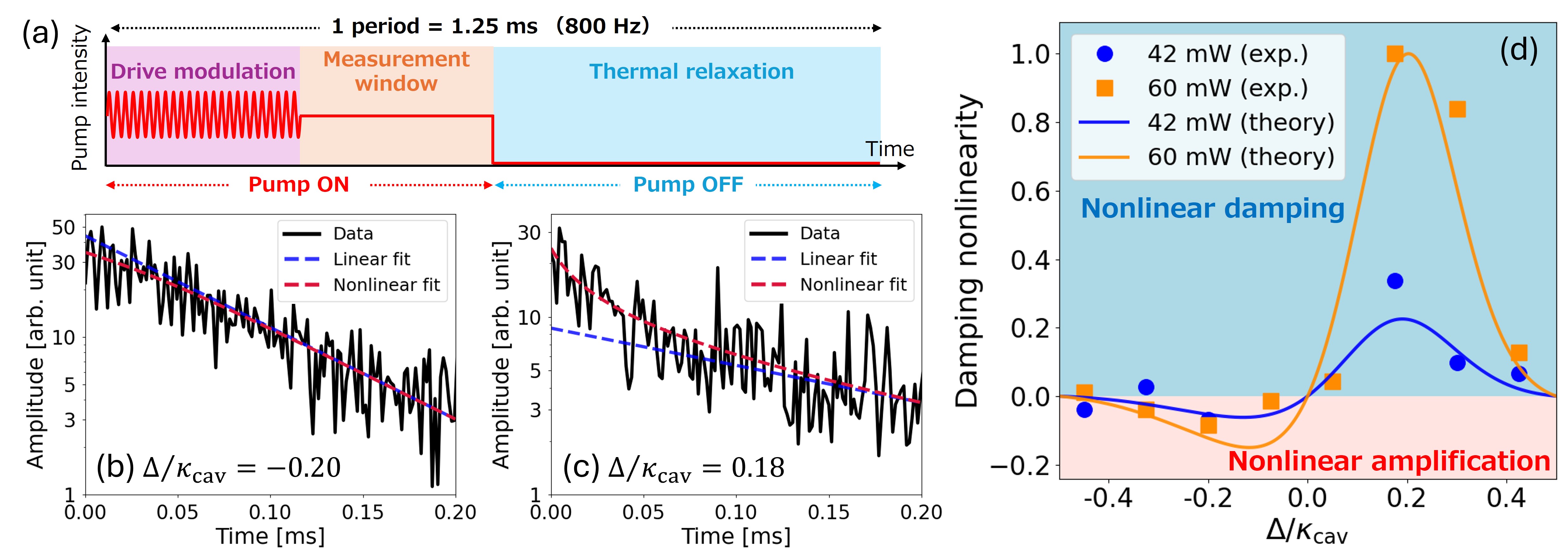}
\caption{
(a) Schematic of the ring-down measurement using pulsed optical excitation to suppress photothermal effects. The pump laser is intensity-modulated into an 800 Hz square-wave pulse train while preserving the peak power; one period, including a pump-off interval for thermal relaxation, is shown. During each pump-on interval, additional modulation first drives the mechanical mode and is then switched off for measurement, while the pump remains on to induce nonlinear mechanical damping.
(b, c) Representative ring-down traces exhibiting negative nonlinear damping (nonlinear amplification) in the red-detuned regime (b) and positive nonlinear damping in the blue-detuned regime (c).
(d) Detuning dependence of the damping nonlinearity $R_{\mathrm{nl}}=\beta_{\mathrm{opt}}x_0^2/(4\Gamma_{\mathrm{eff}})$ for different pump powers. The corresponding ring-down solution is $x(t)=x_0\exp(-\Gamma_{\mathrm{eff}}t/2)/
\sqrt{1+R_{\mathrm{nl}}[1-\exp(-\Gamma_{\mathrm{eff}}t)]}$.
Symbols denote experimental values extracted by fitting the experimental traces; solid curves are calculated using $\beta_{\mathrm{opt}}$ from Fig. \ref{fig:1}(b), the steady-state amplitude $x_0$, and the effective linear damping $\Gamma_{\mathrm{eff}}$, with the radiation-pressure drive taken to be proportional to the intracavity photon number $n_{\mathrm{cav}}(\Delta)$. The plotted values are normalized by the maximum $R_{\mathrm{nl}}$ at an input power of 60 mW. Positive and negative values correspond to positive and negative nonlinear damping, respectively.
}
\label{fig:3}
\end{figure*}

To assess whether the damping enhancement observed in Fig.~\ref{fig:2} originates from optically induced nonlinear damping, we measured the ring-down responses in both the blue- and red-detuned regimes. At the relatively high pump power used here ($\sim$50 mW), however, optical absorption heats the resonator and shifts the cavity resonance through the photothermal effect, making stable access to the red-detuned regime difficult~\cite{carmon2004dynamical}.
To access this regime, we used an intensity modulator to generate a square-wave pump pulse train, thereby reducing the average optical power while maintaining the same peak power [Fig.~\ref{fig:3}(a)]. The ring-down traces in Figs.~\ref{fig:3}(b) and (c) reveal positive nonlinear damping in the blue-detuned regime and negative nonlinear damping (nonlinear amplification) in the red-detuned regime, respectively. 

For quantitative comparison, we define the dimensionless damping nonlinearity as $R_{\mathrm{nl}}=\beta_{\mathrm{opt}}x_0^2/(4\Gamma_{\mathrm{eff}})$, which compares the nonlinear damping at the initial ring-down amplitude $x_0$ with the effective linear damping. The extracted damping nonlinearity follows the theoretically predicted detuning dependence [Fig.~\ref{fig:3}(d)]. Unlike $\beta_{\mathrm{opt}}$ itself, shown in Fig.~\ref{fig:1}(b), $R_{\mathrm{nl}}$ is asymmetric in detuning because its normalization $\Gamma_{\mathrm{eff}}$ is modified by linear optomechanical backaction. In particular, optomechanical cooling increases $\Gamma_{\mathrm{eff}}$ in the red-detuned regime, suppressing $|R_{\mathrm{nl}}|$ and making the signature of negative nonlinear damping less pronounced in the ring-down response. 
Together, the observed sign reversal and quantitative agreement across both detuning regimes demonstrate that the nonlinear damping is optically tunable in both magnitude and sign and provide strong evidence that the damping enhancement originates from delayed optomechanical backaction rather than from intrinsic geometric nonlinearities or residual photothermal effects~\cite{atalaya2016nonlinear,vacher2005anharmonic}. Consistently, the geometrical nonlinearity $\alpha_{\mathrm{geo}}$ is estimated to be much smaller than the optomechanical one, $\alpha_{\mathrm{geo}} \ll \alpha_{\mathrm{opt}}$. As a consequence of this detuning dependence, nonlinear damping terms can be selectively enhanced
or suppressed by tuning the laser detuning, which
enables engineered mechanical nonlinearities through optical
control.

\begin{figure}[t]
\centering
\includegraphics[width=\linewidth]{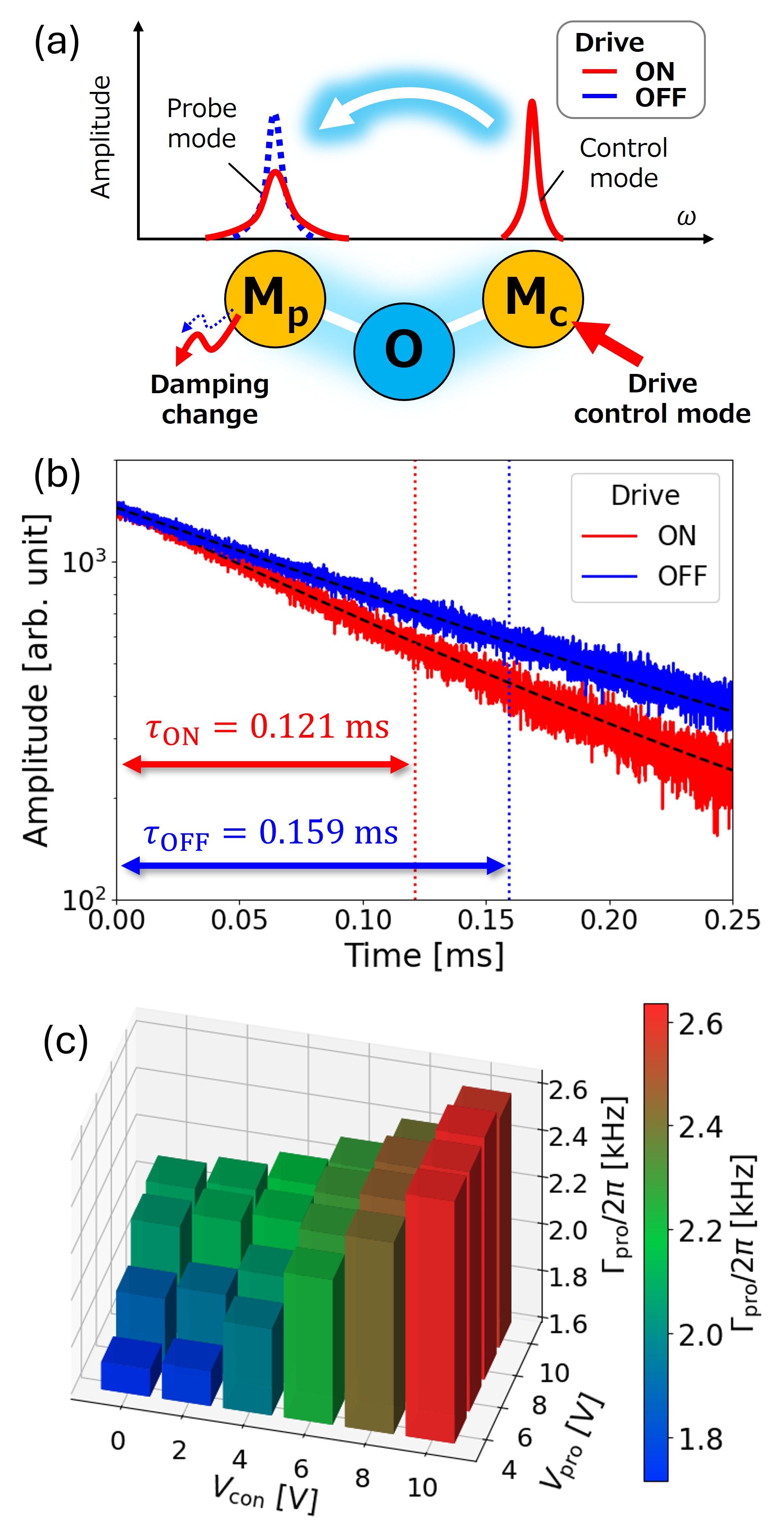}
\caption{
(a) Concept of cross-nonlinear damping in a multimode optomechanical system, where a single optical mode ($\mathrm{O}$) is coupled to two mechanical modes: a probe mode ($\mathrm{M}_{\mathrm{p}}$) and a control mode ($\mathrm{M}_{\mathrm{c}}$). The probe and control modes were resonantly driven at frequencies $f_{\mathrm{pro}} = 48.0$ MHz and $f_{\mathrm{con}} = 48.5$ MHz, respectively.
(b) Ring-down measurements of the probe mode with the control mode strongly driven (ON, red) or undriven (OFF, blue) showing an approximate 0.6 kHz increase in damping. The data were obtained with drive voltages $V_{\mathrm{pro}} = 6.0$ V and $V_{\mathrm{con}} = 10.0$ V, where $V_{\mathrm{pro}}$ and $V_{\mathrm{con}}$ denote the drive voltages for the probe and control modes, respectively.
(c) Increase in probe-mode damping rate $\Gamma_{\mathrm{pro}}/2\pi$ as a function of drive voltages reveals both self-nonlinear damping (dependence on $V_{\mathrm{pro}}$) and cross-nonlinear damping (dependence on $V_{\mathrm{con}}$).
The damping rates are extracted from linear fits to the ring-down traces.
}
\label{fig:4}
\end{figure}

\textit{Cross-nonlinear damping between two mechanical modes - }
Because the nonlinear damping is mediated by the common optical cavity field, multiple mechanical modes coupled to the same optical mode as illustrated in Fig.~\ref{fig:2}(a) can be interconnected through nonlinear mechanical damping. 
Thus, when the amplitude of one mechanical mode increases, the other mode can be strongly damped. This effect is referred to as cross-nonlinear damping [see Fig.~\ref{fig:4}(a)]. Here, we observed cross-nonlinear mechanical damping between two RBMs with resonance frequencies of 48.0 MHz (probe mode) and 48.5 MHz (control mode). We tuned the optical mode so that it couples simultaneously to both RBMs. Figure~\ref{fig:4}(b) shows temporal ring-down traces of the probe mode with (red) and without (blue) excitation of the control mode. The decay time changed from 0.159~ms (2.00~kHz, without excitation) to 0.121~ms (2.62~kHz, with excitation), which constitutes clear evidence of cross-nonlinear mechanical damping. Moreover, we measured the mechanical damping rate of the probe mode, denoted by $\Gamma_{\mathrm{pro}}$, while finely varying the excitation amplitudes of the control and probe modes [see Fig.~\ref{fig:4}(c)]. Damping rate $\Gamma_{\mathrm{pro}}/2\pi$ increased gradually to 2.62~kHz as we increased the excitation of the control mode (via cross-nonlinear damping) and the probe mode itself (via the self-nonlinear damping reported in Fig.~\ref{fig:2}). Based on this, the cross-nonlinear damping coefficient, $\beta_\mathrm{cross}/2\pi$, was estimated to be $2.8\times10^{24}~\mathrm{Hz}/\mathrm{m}^2$, showing agreement in order of magnitude with the theoretical estimate.

\textit{Discussion - }We developed the theoretical framework for nonlinear damping induced by delayed optomechanical backaction, and experimentally demonstrated optically tunable nonlinear damping in an optomechanical resonator. These results reveal the mechanism that generates nonlinear mechanical damping and they show that its strength can be tuned by laser detuning.
The optically tunable nonlinear mechanical damping offers three potential advantages.
(1) The strength of the nonlinear mechanical damping can be tuned on demand and is independent of material composition and geometric design \cite{atalaya2016nonlinear}. This capability enables systematic exploration of tunable nonlinear dynamics governed by nonlinear damping, in close analogy to Duffing nonlinearities, which have recently opened access to a variety of topological phenomena \cite{villa2025topological}. 
Furthermore, unlike structurally designed nonlinear damping, the proposed scheme enables temporal control of nonlinear damping through modulation of the laser power or detuning, opening a route to optomechanical Floquet engineering \cite{mercade2021floquet,pelka2022floquet,Asano2025} of dissipation and the associated nonequilibrium dynamics \cite{sinitsyn2009stochastic}.
(2) RBMs can reach frequencies of several gigahertz not only in WGM optomechanical resonators \cite{ding2010high} but also in photonic crystal cavities \cite{eichenfield2009optomechanical}. In contrast to earlier experiments based on flexural modes, optomechanical schemes exploiting RBMs operate at much higher frequencies, providing a strong advantage for fast damping control. Moreover, high-frequency breathing modes are significantly less affected by air viscous damping, thereby simplifying the experimental configuration under ambient conditions. 
(3) Cross-nonlinear mechanical damping can be extended to multiple mechanical modes. 
Because optomechanical systems (including the microbottle resonators studied herein) scale well with the number of mechanical modes, as demonstrated by our previous realization of arrays containing up to 50 coupled microbottle resonators \cite{asano2024fiber}, cross-nonlinear damping can be engineered by designing the optomechanical couplings.

We have shown nonlinear mechanical damping mediated by the optical cavity field. The nonlinear optomechanical coupling not only induces self-nonlinear damping but also allows its magnitude to be tuned via laser detuning. Furthermore, we observed an optically mediated cross-nonlinear damping effect by coupling a single optical mode to two mechanical modes. This advantageous optomechanical scheme will significantly accelerate studies of nonlinear and nonequilibrium dynamics assisted by nonlinear damping.

\begin{acknowledgments}
This work was supported by JSPS KAKENHI grant numbers JP23H05463.
\end{acknowledgments}

\bibliography{bibtex}

\newpage
\onecolumngrid
\clearpage
\begingroup
\centering
{\Large\bfseries Supplemental Material for ``Optically tunable nonlinear mechanical damping in an optomechanical resonator''\par}
\vspace{0.8em}
{\normalsize Hideki Arahari, Motoki Asano, Hiroshi Yamaguchi, Hajime Okamoto\par}
\vspace{0.5em}
{\normalsize Basic Research Laboratories, NTT, Inc., 3-1 Morinosato Wakamiya, Atsugi-shi, Kanagawa 243-0198 Japan\par}
\vspace{0.8em}
\endgroup
\hrule
\vspace{1.0em}

\renewcommand{\theequation}{S\arabic{equation}}
\renewcommand{\thefigure}{S\arabic{figure}}
\setcounter{equation}{0}
\setcounter{figure}{0}

\section{Memory effects in nonlinear optomechanical coupling}
We consider a single optical cavity mode that is simultaneously coupled via radiation pressure to multiple mechanical modes. The equations of motion for this optomechanical system are
\begin{gather}
    \dot{a}(t) = \left[i\Delta - \frac{\kappa_{\mathrm{cav}}}{2}\right] a(t)
           + i\sum_j g_{j} q_j(t) a(t) + \sqrt{\kappa_{\mathrm{in}}}\, a_{\mathrm{in}}(t), \\
    \ddot{q}_j(t) + \Gamma_j\dot{q}_j(t) + \Omega_j^2 q_j(t)
              = 2g_{j} \Omega_j |a(t)|^{2}.    
\end{gather}
Here, $a(t)$ denotes the amplitude of the intracavity optical mode driven by input field $a_{\mathrm{in}}(t)$. The laser detuning is defined as $\Delta = \omega_{\mathrm{L}} - \omega_{\mathrm{cav}}$, where $\omega_{\mathrm{L}}$ is the angular frequency of the input laser and $\omega_{\mathrm{cav}}$ is the cavity resonance frequency. Parameter $\kappa_{\mathrm{cav}}$ is the total cavity decay rate with external coupling rate $\kappa_{\mathrm{in}}$. The mechanical displacement normalized by zero-point fluctuation amplitude, $q_j(t) (\equiv x_j/x_{\mathrm{zpf},j})$, has resonance frequency $\Omega_{j}$, damping rate $\Gamma_{j}$, and effective mass $m_{\mathrm{eff},j}$. Linearized optomechanical coupling strength $G_j$ and single-photon optomechanical coupling strength $g_j$ are given by $G_j = \partial \omega_{\mathrm{cav}}/\partial x_j$ and $g_{j}= G_j x_{\mathrm{zpf},j}$, respectively. The zero-point fluctuation amplitude is $x_{\mathrm{zpf},j}= \sqrt{\hbar/(2m_{\mathrm{eff},j}\Omega_{j})}$. In the following, we assume a continuous-wave drive, so $a_{\mathrm{in}}(t)=a_{\mathrm{in}}= \mathrm{const.}$ In addition, we introduce rotating-frame transformation $a(t) = \tilde{a}(t)e^{\left(i\Delta - \frac{\kappa_{\mathrm{cav}}}{2}\right)t} = \tilde{a}(t)e^{\lambda_0 t}$, which factorizes out the fast time-evolution of the photon mode
\begin{eqnarray}
    \dot{\tilde{a}}(t) &=& i \sum_j g_j q_j\left(t\right)\tilde{a}\left(t\right) + \sqrt{\kappa_{\mathrm{in}}}a_{\mathrm{in}}e^{-\lambda_0 t},
\end{eqnarray}
where we introduce $\lambda_0=i \Delta-\kappa_{\mathrm{cav}} / 2$ to 
simply express the equations.

To introduce a description that explicitly accounts for the delayed optical response with respect to the mechanical motion, we treat the photon mode using a formal integral solution that incorporates the history (memory) of the mechanical displacement. 
Integrating this equation from $s= -\infty$ to $t$, we obtain
\begin{eqnarray}
    \tilde{a}(t) = \int_{-\infty}^{t} \dot{\tilde{a}}(s)\,ds 
    &=& \sqrt{\kappa_{\mathrm{in}}}a_{\mathrm{in}}\int_{-\infty}^{t}e^{-\lambda_0 s} ds 
    + i \sum_j g_j \int_{-\infty}^{t} q_j(s)\tilde{a}(s) ds \nonumber\\
    &=&  \sqrt{\kappa_{\mathrm{in}}}a_{\mathrm{in}} \chi_0(\Delta) e^{-\lambda_0 t}  + i \sum_j g_j \int_{-\infty}^{t} q_j(s)\tilde{a}(s) ds,  \label{eq:a_base}
\end{eqnarray}
where the susceptibility is $\chi_0(\Delta) = \left(\kappa_{\mathrm{cav}}/2-i\Delta \right)^{-1}$. The right-hand side of Eq.~(\ref{eq:a_base}) again depends on $\tilde{a}$. By recursively substituting the formal solution into the right-hand side, higher-order terms are generated. These recursively generated terms give rise to the nonlinear optomechanical coupling. Expanding in powers of $g_j$ yields the first-order optomechanical coupling and higher-order contributions. Substituting the formal solution again into the higher-order term and truncating at the third order in $g_j$, we obtain
\begin{eqnarray}
    \tilde{a}(t) 
    &\approx& \sqrt{\kappa_{\mathrm{in}}}a_{\mathrm{in}} \chi_0(\Delta) e^{-\lambda_0 t} 
    + i \sqrt{\kappa_{\mathrm{in}}}a_{\mathrm{in}} \chi_0(\Delta) \sum_j g_j \int_{-\infty}^{t}q_j(s)e^{-\lambda_0 s}ds  \nonumber\\ 
    &&+ i^2 \sqrt{\kappa_{\mathrm{in}}}a_{\mathrm{in}}\chi_0(\Delta) \sum_{j,k} g_jg_k \int_{-\infty}^{t}\int_{-\infty}^{s} q_j(s)q_k(s^{\prime}) e^{-\lambda_0 s^{\prime}} dsds^{\prime} \nonumber\\
    && +i^3\sqrt{\kappa_{\mathrm{in}}}a_{\mathrm{in}} \chi_0(\Delta)  \sum_{j,k,l} g_jg_kg_l \int_{-\infty}^{t}\int_{-\infty}^{s}\int_{-\infty}^{s^{\prime}} q_j(s)q_k(s^{\prime})q_l(s^{\prime\prime}) e^{-\lambda_0 s^{\prime\prime}} dsds^{\prime}ds^{\prime\prime}. \label{eq:NL_OM}
\end{eqnarray}
In the final expression, the third and fourth terms on the right-hand side correspond to the second- and third-order nonlinear optomechanical coupling terms, respectively. 

If we restrict the analysis to linear (first-order) optomechanical coupling, the terms beyond the second line of Eq.~(\ref{eq:NL_OM}) can be ignored. Introducing a change in variables, $s=t-\tau$ with $\tau\geq 0$, the integral in the second term, $J_1^{(j)}(t)$, becomes
\begin{eqnarray}
   J_{1}^{(j)}(t) &=& \int_{-\infty}^{t}q_j(s)e^{-\lambda_0 s} ds 
   = e^{-\lambda_0 t}\int_0^{\infty} K^{(1)}(\tau) q_j(t-\tau) d \tau, \label{eq:J_1_general}
\end{eqnarray}
where
\begin{eqnarray}
    \lambda_0&=&i \Delta-\kappa_{\mathrm{cav}} / 2,\quad
    K^{(1)}\left(\tau\right) 
    = e^{\lambda_0\tau}.
\end{eqnarray}
Here $K^{(1)}(\tau)$ represents the first-order memory kernel, and Eq.~(\ref{eq:J_1_general}) can be interpreted as a memory integral over delay time $\tau$. This representation allows us to incorporate explicitly how the past mechanical displacement, $q_j(t-\tau)$, contributes to the present optical response.

In an analogous way, we can introduce memory kernels for the second- and third-order optomechanical coupling terms. For the double integral (third term in Eq.~(\ref{eq:NL_OM})) we define $J_2^{(j,k)}(t)$ as
\begin{eqnarray}
    J_{2}^{(j,k)}(t)
    =\int_{-\infty}^{t} q_j(s) \int_{-\infty}^{s} q_k(s^{\prime}) e^{-\lambda_0 s^{\prime}} ds^{\prime} ds,
\end{eqnarray}
where $\lambda_0 \equiv i \Delta-\kappa_{\mathrm{cav}}/2$. Introducing time differences $\tau_1=t-s$ and $\tau_2=s-s^{\prime}$, {\it i.e.}, $s=t-\tau_1$ and $s^{\prime}=t-\tau_1-\tau_2$, and noting that $\tau_1 \geq 0$ and $\tau_2 \geq 0$, the integral can be rewritten as
\begin{eqnarray}
    J_{2}^{(j,k)}(t) 
    &=& e^{-\lambda_0 t} \int_0^{\infty} \int_0^{\infty}  K^{(2)}\left(\tau_1, \tau_2\right) \mathcal{X}_{j,k}(\tau_1, \tau_2) d \tau_1 d \tau_2, \label{eq:J_2_general} \\
    && \mathcal{X}_{j,k}(\tau_1, \tau_2) = q_j\left(t-\tau_1\right) q_k\left(t-\tau_1-\tau_2\right), \\
    && K^{(2)}\left(\tau_1, \tau_2\right) 
    = e^{\lambda_0 \tau_1}  e^{\lambda_0\tau_2}.
\end{eqnarray}
This is referred to as the second-order memory integral.

Finally, triple integral $J_3^{(j,k,l)}(t)$, which corresponds to the integral part of the fourth term in Eq.~(\ref{eq:NL_OM}), can be written as
\begin{eqnarray}
    J_{3}^{(j,k,l)}(t)
    = \int_{-\infty}^{t} q_j(s) \int_{-\infty}^{s} q_k(s^{\prime}) \int_{-\infty}^{s^{\prime}} q_l(s^{\prime\prime}) e^{-\lambda_0 s^{\prime\prime}} ds^{\prime\prime} ds^{\prime} ds.
\end{eqnarray}
Introducing time differences $\tau_1=t-s$, $\tau_2=s-s^{\prime}$, and $\tau_3=s^{\prime}-s^{\prime \prime}$, {\it i.e.}, $s=t-\tau_1$, $s^{\prime}=t-\tau_1-\tau_2$, and $s^{\prime \prime}=t-\tau_1-\tau_2-\tau_3$, and using $\tau_1, \tau_2, \tau_3 \geq 0$, we obtain the following.
\begin{eqnarray}
    J_{3}^{(j,k,l)}(t) &=& e^{-\lambda_0 t} \int_0^{\infty} \int_0^{\infty} \int_0^{\infty} K^{(3)}\left(\tau_1, \tau_2, \tau_3\right) \mathcal{X}_{j,k,l}(\tau_1, \tau_2, \tau_3) d \tau_1 d \tau_2 d \tau_3 \label{eq:J_3_general}\\
    && \mathcal{X}_{j,k,l}(\tau_1, \tau_2, \tau_3) = q_j\left(t-\tau_1\right) q_k\left(t-\tau_1-\tau_2\right) q_l\left(t-\tau_1-\tau_2-\tau_3\right), \\
    && K^{(3)}\left(\tau_1, \tau_2, \tau_3\right)=e^{\lambda_0 \tau_1} e^{\lambda_0 \tau_2} e^{\lambda_0 \tau_3},
\end{eqnarray}
where $K^{(3)}$ is the third-order memory kernel.
Using the definitions of the first-, second-, and third-order optomechanical memory integrals in Eqs.~(\ref{eq:J_1_general}) - (\ref{eq:J_3_general}),
the solution of the photon-field in Eq.~(\ref{eq:NL_OM}), which was transformed back from the rotating frame, can be compactly rewritten as
\footnotesize
\begin{gather}
    a(t) 
    = \sqrt{\kappa_{\mathrm{in}}}a_{\mathrm{in}} \chi_0(\Delta) \left(1 +i \sum_j g_j J_1^{(j)}(t)e^{\lambda_0 t}
    + i^2 \sum_{j,k} g_jg_k J_2^{(j,k)}(t)e^{\lambda_0 t} + i^3 \sum_{j,k,l}g_jg_kg_l J_3^{(j,k,l)}(t)e^{\lambda_0 t} \right).
\end{gather}
\normalsize

Next, we also expand the equation of motion for the mechanical displacement. In the parameter regime of interest, $\kappa_{\mathrm{cav}} \sim \Omega_{j}$ (partly resolved sideband regime), the optomechanical interaction is in the non-adiabatic regime such that the mechanical resonance frequency appears as sidebands in the optical response. Mechanical displacement $q_j(t)$ is treated as a single-frequency oscillation at $\Omega_{j}$ with slowly varying complex amplitude $B_j(t)$ and can be written as
\begin{eqnarray}
    q_j(t)=B_{j}(t) e^{-i \Omega_{j} t} +\mathrm{c.c.} \label{eq:x=be-iwt}
\end{eqnarray}
Then $\dot{q}_j(t)$ and $\ddot{q}_j(t)$ are, respectively, given by
\begin{eqnarray}
    \begin{gathered}
    \dot{q}_j(t)=\left(\dot{B}_{j}-i \Omega_{j} B_{j}\right) e^{-i \Omega_{j} t} + \mathrm{c.c.}, \\
    \ddot{q}_j(t)=\left(\ddot{B}_{j}-2 i \Omega_{j} \dot{B}_{j}-\Omega_{j}^2 B_j\right) e^{-i \Omega_{j} t} + \mathrm{c.c.}
    \end{gathered}
\end{eqnarray}
Collecting terms proportional to $e^{- i \Omega_{j} t}$, we obtain the exact equations of motion for complex amplitudes
\begin{eqnarray}
    \ddot{B}_j - 2 i \Omega_{j} \dot{B}_{j} - \Omega_j^2 B_j
    +\Gamma_{j}\left(\dot{B}_{j} - i \Omega_{j} B_{j}\right) + \Omega_{j}^2B_j(t) = 2g_j\Omega_j|a(t)|^2.
\end{eqnarray}
Assuming that the envelope varies slowly compared to the mechanical oscillation,
$\left|\dot{B}_{j}\right| \ll \Omega_{j}\left|B_{j}\right|$ and $\left|\ddot{B}_{j}\right| \ll \Omega_{j}\left|\dot{B}_{j}\right|$, we ignore $\ddot{B}_{j}$ and $\Gamma_{j} \dot{B}_{j}$ compared with $\Omega_{j} \dot{B}_{j}$. Then the equation of motion for $B_{j}$ reduces to
\begin{eqnarray}
    &&-2 i \Omega_{j} \dot{B}_{j}(t)-i \Gamma_{j} \Omega_{j} B_{j}(t) \approx  2g_j\Omega_j|a(t)|^2 \nonumber\\
    &&\Rightarrow \dot{B}_{j}(t)=- \frac{\Gamma_{j}}{2} B_{j}(t) +ig_j|a(t)|^2 .
\end{eqnarray}
The envelope equation for mode $j$ can be rewritten by introducing complex oscillation amplitude $b_j(t) = B_j(t) e^{-i \Omega_{j} t}$ as
\begin{eqnarray}
    \dot{b}_j(t) = -i \Omega_{j} b_j(t) -\frac{\Gamma_{j}}{2} b_j(t) + i g_j |a(t)|^2 e^{-i\Omega_j t} \label{eq:Motion_eq}.
\end{eqnarray}

We next apply the same procedure to a nonlinear mechanical oscillator with Duffing and nonlinear damping terms, transforming its equation of motion into one for slowly varying envelope $B_j(t)$
\begin{eqnarray}
   \ddot{q}_j+ \left(\Omega_{j}^2 + \sum_{i} \alpha_{ij}x_{\mathrm{zpf},i}^2 q_i^2\right) q_j + \left(\Gamma_{j} + \sum_{i} \beta_{ij}x_{\mathrm{zpf},i}^2 q_i^2\right) \dot{q_j} 
   = 2g_j\Omega_j |a(t)|^2.
\end{eqnarray}
Here, $\alpha_{ij}$ and $\beta_{ij}$ denote the Duffing and nonlinear-damping coefficients, respectively. The terms with $i=j$ represent self-nonlinearity, while those with $i\neq j$ correspond to cross-nonlinearity arising from other mechanical modes. The Duffing term can be expanded as
\begin{eqnarray}
     q_i^2 q_j &=& \left(B_i^2 e^{-2 i \Omega_{i} t}+2\left|B_{i}\right|^2 + B^{*2}_{i} e^{2 i \Omega_{i} t}\right) \left( B_{j} e^{-i \Omega_{j} t} + \mathrm{c.c.} \right)\nonumber\\
     &=& B_i^2 B_{j} e^{-2 i \Omega_{i} t}e^{-i \Omega_{j} t} + 2\left|B_{i}\right|^2B_{j} e^{-i \Omega_{j} t} + B^{*2}_{i} B_{j} e^{2 i \Omega_{i} t}e^{-i \Omega_{j} t} 
     + \mathrm{c.c.}  
\end{eqnarray}
When the two mechanical frequencies are nearly equal
\begin{eqnarray}
    \abs{\Omega_i - \Omega_j}/\Omega_i  \ll 1, \quad \abs{\Omega_i - \Omega_j}/\Omega_j \ll 1
\end{eqnarray}   
the slowly varying (positive-frequency) component relevant to self- and cross-Duffing interactions reduces to
\begin{eqnarray}
     \left. q_i^2 q_j \right|_+ &\approx& 3\left|B_{i}\right|^2B_{j} e^{-i \Omega_{j} t}.
\end{eqnarray}

For the nonlinear damping term, we have
\begin{eqnarray}
    q_i^2 \dot{q}_j
    &=& \left(B_i^2 e^{-2 i \Omega_{i} t}+2\left|B_{i}\right|^2 + B^{*2}_{i} e^{2 i \Omega_{i} t}\right) \left[ \left(\dot{B}_{j}-i \Omega_{j} B_{j}\right) e^{-i \Omega_{j} t}+ \mathrm{c.c.} \right].
\end{eqnarray}
Extracting the positive-frequency component yields
\begin{eqnarray}
    \left.q_i^2 \dot{q}_j\right|_{+} &=& 2\left|B_{i}\right|^2 \left(\dot{B}_{j}-i \Omega_{j} B_{j}\right) e^{-i \Omega_{j} t}
    + B_i^2 e^{-2 i \Omega_{i} t} \left(\dot{B}_{j}^*+i \Omega_{j} B_{j}^*\right) e^{i \Omega_{j} t}.
\end{eqnarray}
Under the same near-degeneracy condition as Duffing terms
\begin{eqnarray}
    \abs{\Omega_i - \Omega_j}/\Omega_i  \ll 1, \quad \abs{\Omega_i - \Omega_j}/\Omega_j \ll 1
\end{eqnarray}   
dominant slowly varying component ($\left|\dot{B}_{j}\right| \ll \Omega_{j}\left|B_{j}\right|$) becomes
\begin{eqnarray}
    \left.q_i^2 \dot{q}_j\right|_{+}
    &\approx& -i\Omega_{j}\left|B_{i}\right|^2 B_{j} e^{-i \Omega_{j} t}.
\end{eqnarray}
Since the system satisfies the near-degenerate condition for the mechanical frequencies, the nonlinear envelope equation for mode $j$ can be rewritten by introducing complex oscillation amplitude $b_j(t) = B_j(t) e^{-i \Omega_{j} t}$ as
\begin{eqnarray}
  \dot{b}_{j}
    &=&-\left(\frac{\Gamma_{j}}{2}+\sum_{i} \frac{\beta_{ij}}{2}x_{\mathrm{zpf},i}^2\left|b_{i}\right|^2\right) b_{j}
    -i \left(\Omega_j + \sum_{i} \frac{3}{2 \Omega_{j}} \alpha_{ij}x_{\mathrm{zpf},i}^2 \left|b_{i}\right|^2 \right) b_{j} \nonumber\\
    && + i g_j |a(t)|^2 e^{-i \Omega_j t}. \label{eq:Motion_NL_eq_+_Nodim}
\end{eqnarray}

\newpage

\section{Optically induced mechanical nonlinearity in a single mechanical mode}
For clarity, we first analyze the case of a single mechanical mode and derive the higher-order optomechanical interaction terms analytically, thereby demonstrating how optical fields induce mechanical nonlinearities. To focus on a single mechanical oscillator, we rewrite the notation used in the previous section as
\begin{eqnarray}
     m_{\mathrm{eff},j} \rightarrow m_{\mathrm{eff}},\quad
     q_j \rightarrow q,\quad
    \Omega_j \rightarrow \Omega_{\mathrm{m}},\quad 
    \Gamma_j \rightarrow \Gamma_{\mathrm{m}}, \quad
    B_j \rightarrow B,\quad
    b_j \rightarrow b
\end{eqnarray}
and similarly remove mode indices from other related quantities so that no confusion arises.

\subsection{First-order optomechanical coupling}
We now focus on the first-order optomechanical coupling. If the optical response has no delay, {\it i.e.}, $K^{(1)}(\tau)$ does not depend on time, the kernel reduces to delta function $K^{(1)}(\tau)=\delta(\tau)$ and the integral yields
\begin{eqnarray}
    \int_0^{\infty} K^{(1)}(\tau) q(t-\tau) d \tau 
    \simeq q(t)\int_0^{\infty} K^{(1)}(\tau) d \tau = \chi_0 q(t).
\end{eqnarray}
In this case, $\tau=0$ and there is no response delay in the optomechanical coupling. Such a situation corresponds to the bad-cavity limit ($\kappa_{\mathrm{cav}} \gg \Omega_{\mathrm{m}}$, adiabatic regime), where the optical mode responds instantaneously compared to the mechanical motion. In the following, we are interested in memory effects where the past mechanical motion influences the future optical response.

Substituting the single-mode ansatz of Eq.~(\ref{eq:x=be-iwt}) into Eq.~(\ref{eq:J_1_general}) and assuming that $B(t)$ vary slowly on the mechanical timescale ($B(t-\tau)\simeq B(t)$), we obtain
\begin{eqnarray}
   J_{1}(t) e^{\lambda_0 t} &=& \int_0^{\infty} K^{(1)}(\tau) q(t-\tau) d \tau 
   \simeq \int_0^{\infty} K^{(1)}(\tau) \left(B(t) e^{-i \Omega_{\mathrm{m}} (t-\tau)} + \text{c.c.} \right) d \tau \nonumber\\
   &=& B(t) e^{-i \Omega_{\mathrm{m}} t}\int_0^{\infty} K^{(1)}(\tau) e^{i \Omega_{\mathrm{m}} \tau}d \tau + B^*(t) e^{i \Omega_{\mathrm{m}} t}\int_0^{\infty} K^{(1)}(\tau)e^{-i \Omega_{\mathrm{m}} \tau}d \tau \nonumber\\
   &=& \chi^{(1)}_{+} b(t) + \chi^{(1)}_{-} b^*(t), \label{eq:J_1}
\end{eqnarray}
where the memory integrals can be evaluated as
\begin{eqnarray}
    \int_0^{\infty} K^{(1)}e^{\pm i \Omega_{\mathrm{m}} \tau} d \tau 
    =  \int_0^{\infty}e^{\left(i(\Delta\pm\Omega_{\mathrm{m}}) - \kappa_{\mathrm{cav}}/2\right) \tau} d \tau
    = \left[ \frac{e^{\lambda_{\pm} \tau}}{\lambda_{\pm} }\right]_0^{\infty}
    =\frac{1}{(-\lambda_{\pm})} 
    \equiv \chi^{(1)}_{\pm},
\end{eqnarray}
with $\lambda_{\pm} = i(\Delta\pm\Omega_{\mathrm{m}}) - \kappa_{\mathrm{cav}}/2$. Thus the integral yields the first-order susceptibility of the optomechanical coupling. In the following, we denote these susceptibilities by $\chi^{(1)}_{\pm}$.
\newpage

\subsection{Second-order optomechanical coupling}
We next evaluate the second-order optomechanical coupling term. Substituting the single-mode ansatz of Eq.~(\ref{eq:x=be-iwt}) into Eq.~(\ref{eq:J_2_general}) and again assuming that $B(t-\tau)\simeq B(t)$, we obtain
\begin{eqnarray}
    J_{2}(t) e^{\lambda t}
    &=& \int_0^{\infty} \int_0^{\infty}  K^{(2)}\left(\tau_1, \tau_2\right)q\left(t-\tau_1\right) q\left(t-\tau_1-\tau_2\right) d \tau_1 d \tau_2 \nonumber\\
    &\simeq& \int_0^{\infty} \int_0^{\infty}  K^{(2)}\left(\tau_1, \tau_2\right)\left(B(t) e^{-i \Omega_{\mathrm{m}} (t-\tau_1)} + \text{c.c.} \right) \left(B(t) e^{-i \Omega_{\mathrm{m}} (t-\tau_1-\tau_2)} + \text{c.c.} \right) d \tau_1 d \tau_2 \nonumber\\
    &=& B^2(t) e^{-2i \Omega_{\mathrm{m}} t} \int_0^{\infty} \int_0^{\infty}  K^{(2)}\left(\tau_1, \tau_2\right) e^{2i \Omega_{\mathrm{m}} \tau_1}e^{i \Omega_{\mathrm{m}} \tau_2} d \tau_1 d \tau_2\nonumber\\ 
    &&+ B(t)B^*(t) \int_0^{\infty} \int_0^{\infty}  K^{(2)}\left(\tau_1, \tau_2\right) \left( e^{-i \Omega_{\mathrm{m}} \tau_2} + e^{i \Omega_{\mathrm{m}} \tau_2} \right)d \tau_1 d \tau_2 \nonumber\\
    &&+ B^{*2}(t) e^{2i \Omega_{\mathrm{m}}t} \int_0^{\infty} \int_0^{\infty}  K^{(2)}\left(\tau_1, \tau_2\right) e^{-2i \Omega_{\mathrm{m}}\tau_1} e^{-i \Omega_{\mathrm{m}} \tau_2} d \tau_1 d \tau_2\nonumber\\
    &=& \chi^{(1)}_{+2} \chi^{(1)}_{+} B^2(t) e^{-2i \Omega_{\mathrm{m}} t}
    + \chi^{(1)}_{0}\left(\chi^{(1)}_{-} + \chi^{(1)}_{+} \right) B(t)B^*(t) 
    + \chi^{(1)}_{-2} \chi^{(1)}_{-} B^{*2}(t) e^{2i \Omega_{\mathrm{m}}t}, \\
    &=& \chi^{(1)}_{+2} \chi^{(1)}_{+} b^2(t)
    + \chi^{(1)}_{0}\left(\chi^{(1)}_{-} + \chi^{(1)}_{+} \right) b(t)b^*(t) 
    + \chi^{(1)}_{-2} \chi^{(1)}_{-} b^{*2}(t), \label{eq:J_2}
\end{eqnarray}
where we use $\lambda_{\pm n} = \lambda \pm i n \Omega_{\mathrm{m}} = i(\Delta\pm n \Omega_{\mathrm{m}}) - \kappa_{\mathrm{cav}}/2$ and
\begin{eqnarray}
    \int_0^{\infty}\int_0^{\infty} K^{(2)}e^{\pm n i \Omega_{\mathrm{m}} \tau_1} e^{\pm m i \Omega_{\mathrm{m}} \tau_2} d \tau_1 d \tau_2 
    &=&  \int_0^{\infty}\int_0^{\infty}e^{\left(\lambda \pm n i \Omega_{\mathrm{m}} \right) \tau_1}  e^{\left(\lambda \pm m i \Omega_{\mathrm{m}} \right)\tau_2} d \tau_1 d \tau_2 \nonumber\\
    &=& \left[ \frac{e^{\left(\lambda \pm n i \Omega_{\mathrm{m}}\right) \tau_1}}{\lambda \pm n i  \Omega_{\mathrm{m}}}\right]_0^{\infty} \left[ \frac{e^{\left(\lambda \pm m i \Omega_{\mathrm{m}}\right) \tau_2}}{\lambda \pm m i  \Omega_{\mathrm{m}}} \right]_0^{\infty} \nonumber\\
    &=&\frac{1}{(-\lambda_{\pm n})} \frac{1}{\left(-\lambda_{\pm m}\right)}
    \equiv \chi^{(1)}_{\pm n} \chi^{(1)}_{\pm m}.
\end{eqnarray}
Here we expressed the result in terms of first-order optomechanical susceptibilities $\chi^{(1)}$, and $\chi^{(1)}_{0}=(-\lambda_0)^{-1} = \left(\kappa_{\mathrm{cav}}/2 - i\Delta\right)^{-1}$.
\newpage

\subsection{Third-order optomechanical coupling}

Finally, we consider the third-order optomechanical coupling term. Substituting the single-mode ansatz of Eq.~(\ref{eq:x=be-iwt}) into Eq.~(\ref{eq:J_3_general}) and again using $B(t-\tau)\simeq B(t)$, we obtain
\footnotesize
\begin{eqnarray}
    J_{3}(t) e^{\lambda t} &=& \int_0^{\infty} \int_0^{\infty} \int_0^{\infty} K^{(3)}\left(\tau_1, \tau_2, \tau_3\right) q\left(t-\tau_1\right) q\left(t-\tau_1-\tau_2\right) q\left(t-\tau_1-\tau_2-\tau_3\right) d \tau_1 d \tau_2 d \tau_3 \nonumber\\
    &\simeq& \int_0^{\infty} \int_0^{\infty} \int_0^{\infty}  K^{(3)}\left(\tau_1, \tau_2, \tau_3\right) \nonumber\\
    &&\left(B(t) e^{-i \Omega_{\mathrm{m}} (t-\tau_1)} + \text{c.c.} \right) 
    \left(B(t) e^{-i \Omega_{\mathrm{m}} (t-\tau_1-\tau_2)} + \text{c.c.} \right) 
    \left(B(t) e^{-i \Omega_{\mathrm{m}} (t-\tau_1-\tau_2 -\tau_3)} + \text{c.c.} \right) d \tau_1 d \tau_2 d \tau_3 \nonumber\\
    &=& B^3(t) e^{-3i \Omega_{\mathrm{m}} t} \int_0^{\infty} \int_0^{\infty} \int_0^{\infty} K^{(3)}\left(\tau_1, \tau_2, \tau_3\right) e^{3i \Omega_{\mathrm{m}} \tau_1}e^{2i \Omega_{\mathrm{m}} \tau_2}e^{i \Omega_{\mathrm{m}} \tau_3} d \tau_1 d \tau_2 d \tau_3\nonumber\\ 
    &&+ B^2(t)B^*(t) e^{-i \Omega_{\mathrm{m}}t} \nonumber\\
    &&\quad \int_0^{\infty} \int_0^{\infty} \int_0^{\infty} K^{(3)}\left(\tau_1, \tau_2, \tau_3\right) 
    \left( e^{i \Omega_{\mathrm{m}} \tau_1}e^{-i \Omega_{\mathrm{m}} \tau_3} + e^{i \Omega_{\mathrm{m}} \tau_1} e^{i \Omega_{\mathrm{m}} \tau_3} + e^{i \Omega_{\mathrm{m}} \tau_1} e^{2i \Omega_{\mathrm{m}} \tau_2} e^{i \Omega_{\mathrm{m}} \tau_3}\right) 
    d \tau_1 d \tau_2 d \tau_3 \nonumber\\
    &&+ B(t)B^{*2}(t) e^{i \Omega_{\mathrm{m}}t} \nonumber\\
    &&\quad \int_0^{\infty} \int_0^{\infty} \int_0^{\infty} K^{(3)}\left(\tau_1, \tau_2, \tau_3\right) 
    \left( e^{-i \Omega_{\mathrm{m}} \tau_1}e^{-2i \Omega_{\mathrm{m}} \tau_2}e^{-i \Omega_{\mathrm{m}} \tau_3} + e^{-i \Omega_{\mathrm{m}} \tau_1} e^{-i \Omega_{\mathrm{m}} \tau_3} + e^{-i \Omega_{\mathrm{m}} \tau_1} e^{i \Omega_{\mathrm{m}} \tau_3}\right)
    d \tau_1 d \tau_2 d \tau_3 \nonumber\\
    &&+ B^{*3}(t) e^{3i \Omega_{\mathrm{m}}t} \int_0^{\infty} \int_0^{\infty} \int_0^{\infty} K^{(3)}\left(\tau_1, \tau_2, \tau_3\right) e^{-3i \Omega_{\mathrm{m}}\tau_1} e^{-2i \Omega_{\mathrm{m}} \tau_2} e^{-i \Omega_{\mathrm{m}} \tau_3} d \tau_1 d \tau_2 d \tau_3 \nonumber\\
    &=& \chi^{(1)}_{+3} \chi^{(1)}_{+2} \chi^{(1)}_{+} B^3(t) e^{-3i \Omega_{\mathrm{m}} t}
    + \left(\chi^{(1)}_{+}\chi^{(1)}_{0}\chi^{(1)}_{-} + \left(\chi^{(1)}_{+}\right)^2\chi^{(1)}_{0}+ \left(\chi^{(1)}_{+}\right)^2\chi^{(1)}_{+2} \right) B^2(t)B^*(t) e^{-i \Omega_{\mathrm{m}}t} \nonumber\\
    && + \left(\left(\chi^{(1)}_{-}\right)^2\chi^{(1)}_{-2} + \left(\chi^{(1)}_{-}\right)^2\chi^{(1)}_{0} + \chi^{(1)}_{-}\chi^{(1)}_{0}\chi^{(1)}_{+} \right) B(t)B^{*2}(t) e^{i \Omega_{\mathrm{m}}t}
    + \chi^{(1)}_{-3} \chi^{(1)}_{-2} \chi^{(1)}_{-} B^{*3}(t) e^{3i \Omega_{\mathrm{m}}t}, \nonumber\\
    &=& \chi^{(1)}_{+3} \chi^{(1)}_{+2} \chi^{(1)}_{+} b^3(t)
    + \left(\chi^{(1)}_{+}\chi^{(1)}_{0}\chi^{(1)}_{-} + \left(\chi^{(1)}_{+}\right)^2\chi^{(1)}_{0}+ \left(\chi^{(1)}_{+}\right)^2\chi^{(1)}_{+2} \right) \abs{b(t)}^2b(t) \nonumber\\
    && + \left(\left(\chi^{(1)}_{-}\right)^2\chi^{(1)}_{-2} + \left(\chi^{(1)}_{-}\right)^2\chi^{(1)}_{0} + \chi^{(1)}_{-}\chi^{(1)}_{0}\chi^{(1)}_{+} \right) \abs{b(t)}^2b^{*}(t) 
    + \chi^{(1)}_{-3} \chi^{(1)}_{-2} \chi^{(1)}_{-} b^{*3}(t), \label{eq:J_3}
\end{eqnarray}
\normalsize
where the integrals have been evaluated using
\begin{eqnarray}
    &&\iiint_0^{\infty} K^{(3)}e^{\pm l i \Omega_{\mathrm{m}} \tau_1} e^{\pm m i \Omega_{\mathrm{m}} \tau_2} e^{\pm n i \Omega_{\mathrm{m}} \tau_3} d \tau_1 d \tau_2 d \tau_3 \nonumber\\
    &&= \iiint_0^{\infty} e^{\left(\lambda \pm l i \Omega_{\mathrm{m}} \right) \tau_1}  e^{\left(\lambda \pm m i \Omega_{\mathrm{m}} \right)\tau_2} e^{\left(\lambda \pm n i \Omega_{\mathrm{m}} \right)\tau_3} d \tau_1 d \tau_2 d \tau_3 \nonumber\\
    &&= \left[ \frac{e^{\left(\lambda \pm l i \Omega_{\mathrm{m}}\right) \tau_1}}{\lambda \pm l i  \Omega_{\mathrm{m}}}\right]_0^{\infty} \left[ \frac{e^{\left(\lambda \pm m i \Omega_{\mathrm{m}}\right) \tau_2}}{\lambda \pm m i  \Omega_{\mathrm{m}}} \right]_0^{\infty} \left[ \frac{e^{\left(\lambda \pm n i \Omega_{\mathrm{m}}\right) \tau_3}}{\lambda \pm n i  \Omega_{\mathrm{m}}} \right]_0^{\infty} \nonumber\\
    &&=\frac{1}{(-\lambda_{\pm l})} \frac{1}{(-\lambda_{\pm m})} \frac{1}{\left(-\lambda_{\pm n}\right)}
    \equiv \chi^{(1)}_{\pm l} \chi^{(1)}_{\pm m} \chi^{(1)}_{\pm n},
\end{eqnarray}
expressed in terms of first-order susceptibilities $\chi^{(1)}$.
\newpage

\subsection{Mechanical nonlinearity induced by radiation pressure} \label{subsec:1-4}

We now summarize the above to derive the contribution of radiation pressure to the effective mechanical nonlinearity. The photon mode including nonlinear optomechanical coupling [Eq.~(\ref{eq:NL_OM})] can be formally written as
\begin{eqnarray}
    \tilde{a}(t) 
    &=& \sqrt{\kappa_{\mathrm{in}}}a_{\mathrm{in}} \chi_0(\Delta) \left( e^{-\lambda_0 t} +\sum_n \left( i g\right)^n J_{n}(t)\right).
\end{eqnarray}
Here we restrict ourselves to terms up to the third order in $G$. Using Eqs.~(\ref{eq:J_1}), (\ref{eq:J_2}), and (\ref{eq:J_3}), the solution, which is transformed back from the rotating frame, can be written as
\begin{eqnarray}
    a(t) 
    &=& \sqrt{\kappa_{\mathrm{in}}}a_{\mathrm{in}} \chi_0(\Delta) \left(1 +i g J_1(t)e^{\lambda_0 t}
    + (i g)^2 J_2(t)e^{\lambda_0 t} + (i g)^3 J_3(t)e^{\lambda_0 t} \right).
\end{eqnarray}
Using the intracavity photon number in the absence of optomechanical backaction, $n_{\mathrm{cav}}(\Delta)=\kappa_{\mathrm{in}} \abs{a_{\mathrm{in}}}^2\abs{\chi_0(\Delta)}^2$, the intracavity intensity can then be written as
\begin{eqnarray}
    |a(t)|^2 
    &=& n_{\mathrm{cav}}(\Delta)
    \abs{1 +i g J_1(t)e^{\lambda_0 t} + (i g)^2 J_2(t)e^{\lambda_0 t} + (i g)^3 J_3(t)e^{\lambda_0 t} }^2.
\end{eqnarray}
We define
\begin{eqnarray}
    E(t)&=&1+i g \mathcal{A} + (ig)^2 \mathcal{B}+(i g)^3 \mathcal{C} \\
    && \text{where } 
    \mathcal{A}(t) =J_1(t) e^{\lambda_0 t}, \quad 
    \mathcal{B}(t) =J_2(t) e^{\lambda_0 t}, \quad 
    \mathcal{C}(t) =J_3(t) e^{\lambda_0 t},
\end{eqnarray}
and expand $\abs{E(t)}^2$ using
\begin{eqnarray}
    E=1 + i g \mathcal{A} + (ig)^2 \mathcal{B}+ (ig)^3 \mathcal{C}, \quad E^*
    =1-i g \mathcal{A}^*+(-i g)^2 \mathcal{B}^*+(-i g)^3 \mathcal{C}^*.
\end{eqnarray}
When expanding $\abs{E(t)}^2$, the components at frequencies $\pm\Omega_{\mathrm{m}}$ in products $J_1[\pm\Omega_{\mathrm{m}}]$, $J_2[\pm2\Omega_{\mathrm{m}}, 0\Omega_{\mathrm{m}}]$, and $J_3[\pm\Omega_{\mathrm{m}}, \pm3\Omega_{\mathrm{m}}]$ yield contributions $F_{\pm}$ to the radiation pressure force. Focusing on positively rotating component $F_{+}[+\Omega_{\mathrm{m}}]$, the relevant terms are
\begin{eqnarray}
    &J_1[+\Omega_{\mathrm{m}}], \quad
    \left(J_1[-\Omega_{\mathrm{m}}]\right)^*,& \label{eq:J_1_comps}\\
    &J_3[+\Omega_{\mathrm{m}}], \quad
    \left(J_3[-\Omega_{\mathrm{m}}]\right)^*,& \label{eq:J_3_comps}\\
    &J_1[-\Omega_{\mathrm{m}}]\left(J_2[-2\Omega_{\mathrm{m}}]\right)^*, \quad
    \left(J_1[+\Omega_{\mathrm{m}}]\right)^*J_2[+2\Omega_{\mathrm{m}}],& \nonumber\\
    &\qquad J_1[+\Omega_{\mathrm{m}}]\left(J_2[0\Omega_{\mathrm{m}}]\right)^*, \quad 
    \left(J_1[-\Omega_{\mathrm{m}}]\right)^*J_2[0\Omega_{\mathrm{m}}],& \label{eq:J_1_J_2_comps}\\
    &J_2[0\Omega_{\mathrm{m}}]\left(J_3[-\Omega_{\mathrm{m}}]\right)^*, \quad
    \left(J_2[0\Omega_{\mathrm{m}}]\right)^*J_3[+\Omega_{\mathrm{m}}],& \nonumber\\
    &\qquad J_3[+3\Omega_{\mathrm{m}}]\left(J_2[+2\Omega_{\mathrm{m}}]\right)^*, \quad
    \left(J_3[-3\Omega_{\mathrm{m}}]\right)^*J_2[-2\Omega_{\mathrm{m}}],& \label{eq:J_2_J_3_comps}
\end{eqnarray}
{\it i.e.}, twelve terms in total. Since the higher-order terms in $g$ (above the fourth order in intensity) are relatively smaller than the other terms, we ignore them and retain only the eight terms in Eqs.~(\ref{eq:J_1_comps})--(\ref{eq:J_1_J_2_comps}). Note that $\chi^{(1)*}_{\pm} \neq \chi^{(1)}_{\mp}$ in general, so $\abs{E(t)}^2$ can be written as
\begin{eqnarray}
     \abs{E(t)}^2 
     &\simeq& \left( 
     i g J_{1}[+\Omega_{\mathrm{m}}] - i g \left(J_{1}[-\Omega_{\mathrm{m}}]\right)^*
     + \left(i g\right)^3 J_{3}[+\Omega_{\mathrm{m}}]
     + \left(\left(i g\right)^{3} J_{3}[-\Omega_{\mathrm{m}}]\right)^* \right.\nonumber\\
     &&\left.+ \left(i g\right)^3 J_{1}[-\Omega_{\mathrm{m}}]\left(J_{2}[-2\Omega_{\mathrm{m}}]\right)^* 
     - \left( i g\right)^3 \left(J_{1}[+\Omega_{\mathrm{m}}]\right)^*J_{2}[+2\Omega_{\mathrm{m}}] \right.\nonumber\\
     &&\left.+ \left( i g\right)^3 J_{1}[+\Omega_{\mathrm{m}}]\left(J_{2}[0\Omega_{\mathrm{m}}]\right)^*
     - \left(i g\right)^3 \left(J_{1}[-\Omega_{\mathrm{m}}]\right)^* J_{2}[0\Omega_{\mathrm{m}}]
     \right) \nonumber\\
     &=& i g \left(\chi^{(1)}_{+} - \chi^{(1)*}_{-}\right) B(t) \nonumber\\
     &&+ (i g)^3  \left\{\left(\chi^{(1)}_{+}\chi^{(1)}_{0}\chi^{(1)}_{-} + \left(\chi^{(1)}_{+}\right)^2\chi^{(1)}_{0}+ \left(\chi^{(1)}_{+}\right)^2\chi^{(1)}_{+2} \right) \right.\\
     &&\left. \quad - \left(\chi^{(1)}_{-}\chi^{(1)}_{0}\chi^{(1)}_{+} + \left(\chi^{(1)}_{-}\right)^2\chi^{(1)}_{0}+ \left(\chi^{(1)}_{-}\right)^2\chi^{(1)}_{-2} \right)^* \right\} \abs{B}^2 B(t) \nonumber\\
     &&+ (i g)^3 \left(\chi^{(1)}_{-}\chi^{(1)*}_{-2} \chi^{(1)*}_{-} - \chi^{(1)*}_{+}\chi^{(1)}_{+2} \chi^{(1)}_{+} \right) \abs{B}^2 B(t) \nonumber\\
     &&+ (i g)^3 \left\{ \chi^{(1)}_{+} \chi^{(1)*}_{0} \left(\chi^{(1)}_{+} + \chi^{(1)}_{-} \right)^* 
     - \chi^{(1)*}_{-} \chi^{(1)}_{0} \left(\chi^{(1)}_{-} + \chi^{(1)}_{+} \right) \right\}\abs{B}^2 B(t) \\
     &=& - g C_1\left( \Delta \right) B(t) 
    - g^3 \left(  C_2\left(\Delta\right) + C_3\left(\Delta\right) + C_4\left(\Delta\right) \right)\abs{B}^2 B(t),
\end{eqnarray}
where we introduce linear contribution $C_1$ and nonlinear contributions $C_2$, $C_3$, and $C_4$, respectively, as
\begin{eqnarray}
    C_1\left(\Delta\right) &=& -i\left(\chi^{(1)}_{+} - \chi^{(1)*}_{-}\right) 
    =  \frac{1}{(\Delta+\Omega_{\mathrm{m}}) +i \kappa_{\mathrm{cav}}/2} + \frac{1}{(\Delta-\Omega_{\mathrm{m}}) - i \kappa_{\mathrm{cav}}/2}, \\
    C_{2,3,4}\left(\Delta\right) &=& i\left[ S_{2,3,4}^{+} - (S_{2,3,4}^{-})^* \right],\\
    S_{2}^{\pm} &=& \chi^{(1)}_{\pm}\chi^{(1)}_{0}\chi^{(1)}_{\mp} + \left(\chi^{(1)}_{\pm}\right)^2 \chi^{(1)}_{0} + \left(\chi^{(1)}_{\pm}\right)^2\chi^{(1)}_{\pm2}, \\
    S_{3}^{\pm} &=& \chi^{(1)}_{\mp}\chi^{(1)*}_{\mp2} \chi^{(1)*}_{\mp},\\
    S_{4}^{\pm} &=& \chi^{(1)}_{\pm} \chi^{(1)*}_{0} \left(\chi^{(1)}_{\pm} + \chi^{(1)}_{\mp} \right)^*.
\end{eqnarray}
The linear term in $g$ describes the contribution of the optomechanical coupling to the linear modulation of the mechanical motion, while the cubic term in $g$ describes the induced nonlinear modulation. The real and imaginary parts of these coefficients represent contributions to the mechanical resonance frequency and damping rate, respectively. Using the expression of the intracavity intensity,
\begin{eqnarray}
    |a(t)|^2 &=&n_{\mathrm{cav}}(\Delta) |E(t)|^2,
\end{eqnarray}
we can rewrite the third terms of the right-hand side in Eq.~(\ref{eq:Motion_eq}) as
\begin{eqnarray}
    i g |a(t)|^2 e^{-i \Omega_{\mathrm{m}} t} &=&  i  n_{\mathrm{cav}}(\Delta) g |E(t)|^2 e^{-i \Omega_{\mathrm{m}} t} \nonumber\\
    &=&  -i  n_{\mathrm{cav}}(\Delta) g^2 \left( C_1\left( \Delta \right) B(t) 
    + g^2 \left(  C_2\left(\Delta\right) + C_3\left(\Delta\right) + C_4\left(\Delta\right) \right)\abs{B}^2 B(t) \right) e^{-i \Omega_{\mathrm{m}} t}\nonumber\\
    &=&  -i  n_{\mathrm{cav}}(\Delta) g^2 \left( C_1\left( \Delta \right) b(t) 
    + g^2 \left(  C_2\left(\Delta\right) + C_3\left(\Delta\right) + C_4\left(\Delta\right) \right)\abs{b}^2 b(t) \right) .
\end{eqnarray}
From the form of the equation of motion for the nonlinear oscillator [Eq.~(\ref{eq:Motion_NL_eq_+_Nodim})], the linear and nonlinear mechanical modulation terms originating from the optical force can be expressed as follows.
\begin{eqnarray}
    \dot{b}(t) &=& -i \left[\Omega_{\mathrm{eff}} - i\frac{\Gamma_{\mathrm{eff}}}{2}\right]b(t) -i\left(\frac{3\alpha_{\mathrm{opt}}}{2\Omega_{\mathrm{m}}} - i \frac{\beta_{\mathrm{opt}}}{2} \right) x_{\mathrm{zpf}}^2 \left|b\right|^2 b(t) \\
    && \text{where} \quad \Omega_{\mathrm{eff}} = \Omega_{\mathrm{m}} + \Omega_{\mathrm{opt}}(\Delta), \quad \Gamma_{\mathrm{eff}} = \Gamma_{\mathrm{m}}+\Gamma_{\mathrm{opt}}(\Delta)
\end{eqnarray}
Here, we obtain the optomechanically induced linear and nonlinear coefficients
\begin{eqnarray}
    \Omega_{\mathrm{opt}}(\Delta) &=& n_{\mathrm{cav}}(\Delta) g^2 \Re\left[C_1\left(\Delta\right)\right] \label{eq:Omega_opt}\\
    \Gamma_{\mathrm{opt}}(\Delta) &=& -2n_{\mathrm{cav}}(\Delta) g^2 \Im\left[C_1\left(\Delta\right)\right] \label{eq:Gamma_opt}\\
    \alpha_{\mathrm{opt}}(\Delta) &=& \frac{2\Omega_{\mathrm{m}}}{3x_{\mathrm{zpf}}^2} n_{\mathrm{cav}}(\Delta) g^4 \Re\left[ C_2\left(\Delta\right) + C_3\left(\Delta\right) + C_4\left(\Delta\right) \right] \label{eq:alpha_opt}\\ 
    \beta_{\mathrm{opt}}(\Delta) &=& - \frac{2}{x_{\mathrm{zpf}}^2}n_{\mathrm{cav}}(\Delta) g^4 \Im\left[ C_2\left(\Delta\right) + C_3\left(\Delta\right) + C_4\left(\Delta\right) \right] \label{eq:beta_opt},
\end{eqnarray}
where we use $x_{\mathrm{zpf}}= \sqrt{\frac{\hbar}{2m_{\mathrm{eff}}\Omega_{\mathrm{m}}}}$. The ratio of the optomechanically induced nonlinear damping and Duffing coefficients is then given by
\begin{eqnarray}
    \frac{\beta_{\mathrm{opt}}}{\alpha_{\mathrm{opt}}}(\Delta) &=& -\frac{3}{\Omega_{\mathrm{m}}} \frac{\Im\left[ C_2\left(\Delta\right) + C_3\left(\Delta\right) + C_4\left(\Delta\right) \right]}{\Re\left[ C_2\left(\Delta\right) + C_3\left(\Delta\right) + C_4\left(\Delta\right) \right]}.
\end{eqnarray}

Here, the specific parameter values used in the calculations are as follows. The effective mass is taken as $m_{\mathrm{eff}} = 7\times10^{-9}~$kg. This value is estimated by scaling from the effective mass of the radial breathing mode (RBM) of a microbottle resonator reported in Ref.~\citep{Asano2022}. The single-photon optomechanical coupling rate and the zero-point fluctuation amplitude are taken as $g = 2\pi \times 29.6~$Hz and $x_{\mathrm{zpf}} = 5.0\times 10^{-18}~$m, respectively.

\newpage

\section{Optically induced mechanical nonlinearity in multi-mechanical modes}

The intracavity photon mode depends on the displacements of all mechanical modes, as expressed in Eq.~(\ref{eq:NL_OM}). Its formal solution up to the third order in the optomechanical interaction is
\footnotesize
\begin{eqnarray}
    \tilde{a}(t) 
    &=& \sqrt{\kappa_{\mathrm{in}}}a_{\mathrm{in}} \chi_0(\Delta) e^{-\lambda_0 t} 
    + i \sqrt{\kappa_{\mathrm{in}}}a_{\mathrm{in}} \chi_0(\Delta) \sum_j g_j \int_{-\infty}^{t}q_j(s)e^{-\lambda_0 s}ds  \nonumber\\ 
    &&+ i^2 \sqrt{\kappa_{\mathrm{in}}}a_{\mathrm{in}}\chi_0(\Delta) \sum_{j,k} g_jg_k \int_{-\infty}^{t}\int_{-\infty}^{s} q_j(s)q_k(s^{\prime}) e^{-\lambda_0 s^{\prime}} dsds^{\prime} \nonumber\\
    && +i^3\sqrt{\kappa_{\mathrm{in}}}a_{\mathrm{in}} \chi_0(\Delta)  \sum_{j,k,l} g_jg_kg_l \int_{-\infty}^{t}\int_{-\infty}^{s}\int_{-\infty}^{s^{\prime}} q_j(s)q_k(s^{\prime})q_l(s^{\prime\prime}) e^{-\lambda_0 s^{\prime\prime}} dsds^{\prime}ds^{\prime\prime}.
\end{eqnarray}
\normalsize
We assume that the displacement of the $j$-th mechanical mode can be written as a single-frequency vibration, $\Omega_{j}$, with a slowly varying complex envelope, $B_j(t)$,
\begin{eqnarray}
    q_j(t)=B_{j}(t) e^{-i \Omega_{j} t} +\mathrm{c.c.}
\end{eqnarray}
Substituting this expression into each nonlinear optomechanical term and performing the corresponding kernel integrals yields a generalized intracavity field that contains contributions from all mechanical modes. Inserting the resulting photon field into the radiation-pressure term then allows us to derive the optically induced mechanical nonlinearities for the multimode optomechanical system. Although the full expression becomes algebraically cumbersome, its structure parallels the single-mode result as shown in Eqs. (\ref{eq:Omega_opt})-(\ref{eq:beta_opt}). The real and imaginary parts of each optomechanical interaction term, $J_1, J_2,$ and $J_3$, govern the mechanically induced frequency shift and damping modification, respectively. Consequently, the radiation-pressure force acting on the $j$-th mechanical mode takes a form that depends on the amplitudes of all mechanical modes.
\begin{eqnarray}
    F_{\mathrm{opt},j} =\hbar G_j |a(t)|^{2}
    \propto -\left(\Omega_{\mathrm{opt}, j} + \sum_i \alpha_{ij} \abs{b_i}^2\right)b_j 
    + i\left(\Gamma_{\mathrm{opt}, j} + \sum_i \beta_{ij}\abs{b_i}^2 \right) b_j ,\label{eq:F_opt_general}
\end{eqnarray}
where $\Omega_{\mathrm{opt}, j}$ and $\Gamma_{\mathrm{opt}, j}$ denote the optically induced linear frequency shift and damping, respectively. Coefficients $\alpha_{ij}$ and $\beta_{ij}$ quantify the cross-nonlinear coupling between mechanical modes.
As shown in Eq.~(\ref{eq:F_opt_general}), once we consider multiple mechanical modes, the resonance frequency and the damping of a given mode become dependent on the amplitudes of the other mechanical modes. These cross terms give rise to cross-Duffing nonlinearity (frequency shift) and cross-nonlinear damping, respectively. The present formalism enables numerical estimation of all such coefficients.
\newpage

\section{Measurement of thermal motion of mechanical RBM in the microbottle resonator}
\begin{figure}[h]
\centering
\includegraphics[width=0.9\linewidth]{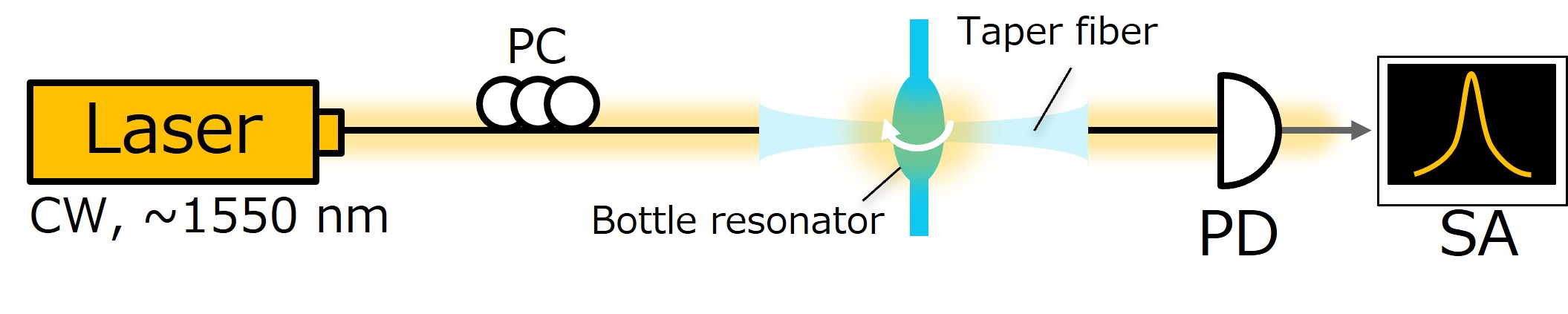}
\caption{
Overview of experimental configuration. PC: Polarization controller, PD: Photodetector, SA: Spectrum analyzer.
\label{fig:S1}}
\end{figure}

We measured the thermal motion of the bottle-shaped optomechanical resonator fabricated on a silica optical fiber. The measurements were conducted under ambient-pressure and room-temperature conditions. Figure \ref{fig:S1} shows a schematic of the experimental configuration used to measure the thermal motion. A laser with a wavelength near 1550 nm is launched into the bottle resonator through a tapered fiber (diameter $\sim$1 \textmu m) with the thermally driven RBM to be transduced into intensity fluctuations of the transmitted optical signal, which are subsequently converted into a vibration spectrum using a spectrum analyzer (SA).
In addition, the thermal-noise spectra obtained by the above procedure are used to calibrate the amplitude of the mechanical displacement. This calibration allows us to convert the measured voltage signals such as those obtained in the ring-down experiment shown in Fig. 2(e) in the main text into absolute displacement units {\it e.g.}, picometers.

\begin{figure}[h]
\centering
\includegraphics[width=0.9\linewidth]{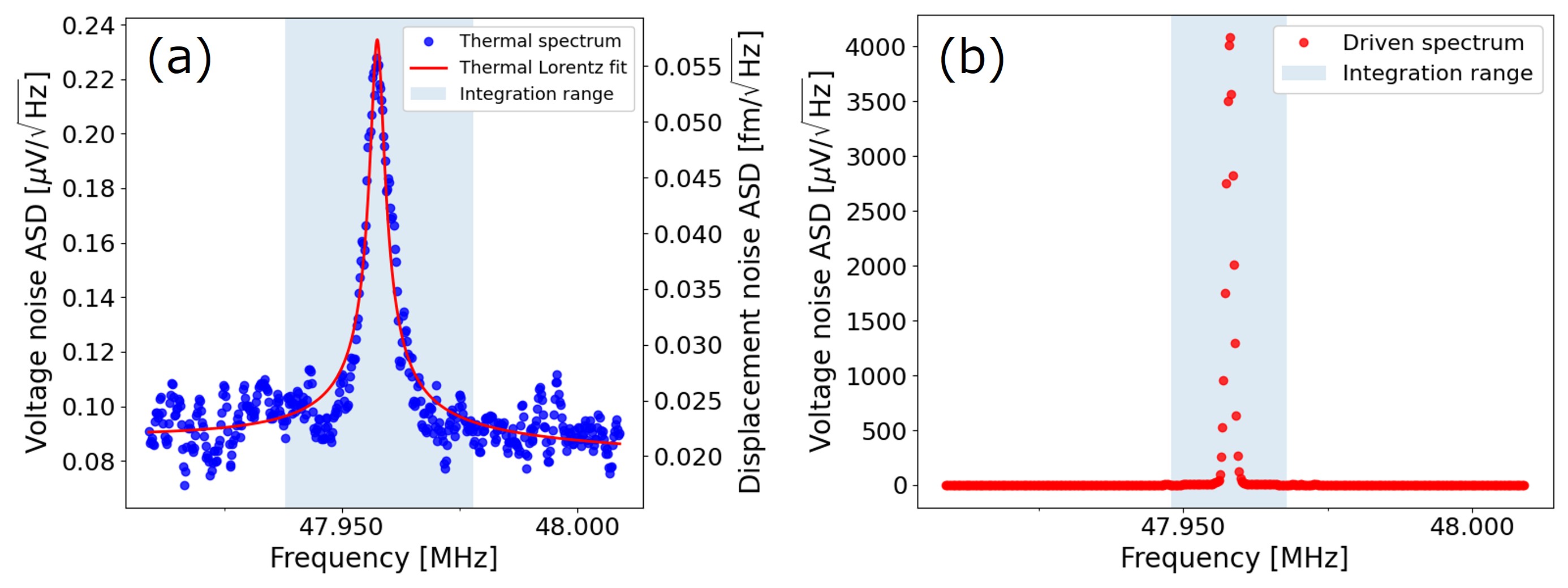}
\caption{
(a)Voltage-noise amplitude spectral density (ASD) of the thermomechanical motion and its Lorentzian fit. The right axis shows the corresponding displacement-noise ASD. (b) Voltage ASD measured with a spectrum analyzer under resonant mechanical excitation at $V_{\mathrm{EOM}}=10$ V. The shaded regions indicate the frequency ranges used for integration. The integrated PSDs are obtained by integrating the squared ASDs over these ranges.
\label{fig:S2}}
\end{figure}

The mechanical amplitude was calibrated using the thermomechanical noise of the resonator. Figure~\ref{fig:S2}(a) shows the thermal noise spectrum measured with a spectrum analyzer, while Fig.~\ref{fig:S2}(b) shows the spectrum obtained under resonant excitation with a drive voltage of 10~V.

The displacement of a mechanical mode at thermal equilibrium is given by the equipartition theorem,
\begin{equation}
    \frac{1}{2}m_{\mathrm{eff}}\Omega_{\mathrm{m}}^2\langle x^2\rangle = \frac{1}{2}k_B T,
\end{equation}
where $m_{\mathrm{eff}}$ is the effective mass, $\Omega_{\mathrm{m}}$ is the mechanical angular resonance frequency, $k_B$ is the Boltzmann constant, and $T$ is the temperature. The root-mean-square (RMS) thermal displacement is therefore
\begin{equation}
    x_{\mathrm{th,rms}} = \sqrt{\frac{k_B T}{m_{\mathrm{eff}}\Omega_m^2}}.
\end{equation}
The area of the displacement power spectral density (PSD) is related to the displacement variance by
\begin{equation}
    \int S_{x,\mathrm{th}}(f)\,df
    = \langle x^2\rangle
    = x_{\mathrm{th,rms}}^2.
\end{equation}
Assuming a linear transduction coefficient between displacement and detected voltage, the integrated voltage PSDs of the thermal and driven spectra can be written as
\begin{eqnarray}
    A_{v,\mathrm{th}} &=& \int S_{v,\mathrm{th}}(f)\,\mathrm{d}f, \\
    A_{v,\mathrm{drv}} &=& \int S_{v,\mathrm{drv}}(f)\,\mathrm{d}f.
\end{eqnarray}
Since both spectra are measured with the same detection chain, the transduction coefficient cancels when taking the ratio of the integrated PSDs. The RMS amplitude under resonant excitation is therefore obtained as
\begin{equation}
    x_{\mathrm{drv,rms}} = x_{\mathrm{th,rms}}
    \sqrt{
    \frac{A_{v,\mathrm{drv}}}
         {A_{v,\mathrm{th}}}
    }.
\end{equation}
For a sinusoidal oscillation, the peak amplitude is related to the RMS amplitude by
\begin{equation}
    x_{\mathrm{drv,peak}} = \sqrt{2}\, x_{\mathrm{drv,rms}}.
\end{equation}
Combining the above expressions yields
\begin{equation}
    x_{\mathrm{drv,peak}} = \sqrt{2} \sqrt{\frac{k_B T}{m_{\mathrm{eff}}\Omega_m^2}} \sqrt{
    \frac{A_{v,\mathrm{drv}}}
         {A_{v,\mathrm{th}}}
    }.
\end{equation}
Using the measured spectra shown in Fig.~\ref{fig:S2} and the independently determined effective mass ($m_{\mathrm{eff}}=7\times 10^{-9}~$kg) and the resonator temperature of $T=300~$K, we obtained the peak mechanical oscillation amplitude under a drive voltage of 10~V as
\begin{equation}
    x_{\mathrm{drv,peak}} = 28.67~\mathrm{pm}.
\end{equation}
This calibration factor was subsequently used to convert the measured voltage amplitudes into physical displacement amplitudes throughout this work.

\newpage

\section{Experimental Configuration for Probing Optically Induced Mechanical Nonlinearity}
In this section, we describe the experimental configuration used to measure the driven mechanical frequency response and ring-down dynamics of a single mechanical mode, as presented in Figs.~2 and 3 in the main text.
As illustrated in Fig. \ref{fig:S1}, lasers are injected into the bottle resonator through evanescent coupling by setting a tapered fiber in contact with the bottle surface. For the excitation and detection of mechanical motion, we prepare two lasers with slightly different wavelengths: a pump laser ($\sim$1550 nm) for driving the mechanical RBM, and a probe laser ($\sim$1520 nm) for monitoring it. The pump laser is intensity-modulated using an electro-optic modulator, with a sinusoidal voltage of amplitude $V$ and angular frequency $\omega_{\mathrm{mod}}$ generated by an arbitrary function generator. In this way, we periodically modulate the radiation pressure force, and resonantly drive the mechanical RBM by driving the modulation near the mechanical resonance ($\omega_{\mathrm{mod}}\sim\Omega_{\mathrm{m}}$). We also tune the modulation depth of the pump intensity and, hence the excitation strength, by the applied voltage. The transmitted pump and probe lasers pass through the cavity, after which the pump is filtered out by a bandpass filter (BPF), and the remaining probe is detected by a lock-in amplifier for mechanical readout.

\begin{figure}
\centering
\includegraphics[width=0.9\linewidth]{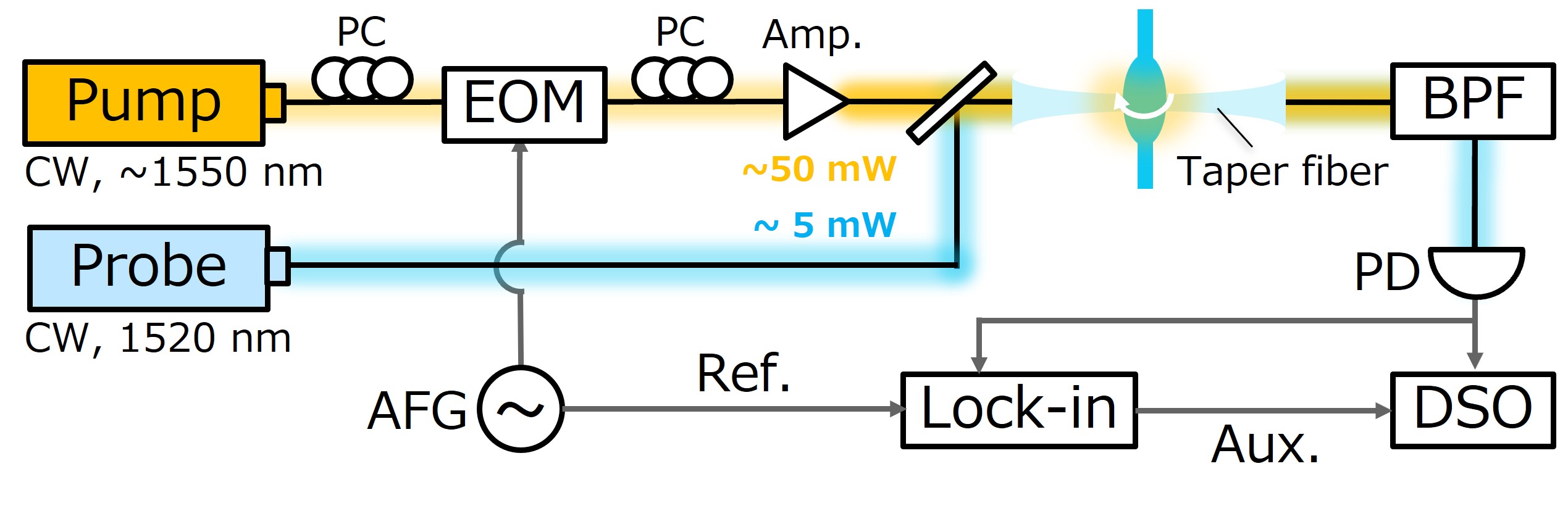}
\caption{
Experimental configuration. EOM: Electro-optic modulator, AFG: Arbitrary function generator, BPF: Bandpass filter, DSO: Digital sampling oscilloscope, PC: Polarization controller, PD: Photodetector.
\label{fig:S3}}
\end{figure}

To investigate the nonlinear mechanical behavior induced by optomechanical coupling, we inject a strong optical field into the cavity so that the radiation-pressure force acquires a nonlinear dependence on the mechanical displacement. As the intracavity photon number increases, higher-order components of the optical force become significant giving rise to optically induced Duffing nonlinearity and nonlinear damping. In the experiments, the pump beam was amplified to an optical power of approximately 50~mW before entering the cavity [Fig. \ref{fig:S3}], enabling access to the nonlinear regime. In the nonadiabatic regime, where the cavity response time is comparable to the mechanical oscillation period, the delayed optical backaction produces an effective damping force that depends on the vibrational amplitude.

Since the modulation frequency is much higher than the mechanical damping rate ($\omega_{\mathrm{mod}} \gg \Gamma_{\mathrm{m}}$), the mechanical oscillator responds to the time-averaged optical force. Taking into account the nonadiabatic nonlinearity of the optical force, the equation of motion for the mechanical oscillator can be written as follows.
\begin{eqnarray}
    \ddot{x} + \left(\Gamma_{\mathrm{m}} + \beta_{\mathrm{opt}} x^{2}\right) \dot{x} &+& \left(\Omega_{\mathrm{m}}^{2} + \alpha_{\mathrm{opt}}x^{2}\right) x 
    = \frac{F_{\mathrm{d}}}{m_{\mathrm{eff}}}\, \cos(\omega_{\mathrm{mod}} t),
\end{eqnarray}
where $\Gamma_{\mathrm{m}}$ and $\Omega_{\mathrm{m}}$, respectively, denote the mechanical damping and resonance frequency including the modulation induced by the linear optomechanical coupling, and $F_{\mathrm{d}}$ represents the amplitude of the driving force induced by the optical force. 
\newpage

\section{Theoretical Estimation of Mechanical Geometric Nonlinearity}
Here we discuss the derivation of Duffing nonlinearity $\alpha_G$ originating from the geometric deformation for the fundamental RBM in the microbottle structure given by
\begin{align}
u_r(r,z)=&- k_r J_1 (k_r r)H_0(\sqrt{k_0\beta}z)\exp\left[-\frac{k_0\beta}{2}z^2\right],\\
u_z(r,z)=&-2 k_0\beta z J_0 (k_r r)\exp\left[-\frac{k_0\beta}{2}z^2\right],
\end{align}
where $k_r=1.84/R_0$ is the main radius of the microbottle, $k_0=\sqrt{k_r^2+\beta^2}+\beta$ is the curvature of $\beta$ parameterizing the shape of the microbottle, and $r(z)=\sqrt{1+\beta^2 z^2}R_0$.  Terms $J_n(x)$ and $H_n(x)$ denote the $n$th order of the Bessel function and Hermite polynomial, respectively. 

The mechanical strain generalized by the structural deformation can be given by the Green-Lagrange strain, $E=\frac{1}{2}(FF^t-I)$, where $F$ shows the deformation tensor given by
\begin{align}
    F=\left(\begin{array}{ccc}
    1+u_{r,r}&0&u_{r,z}\\
    0&1+u_r/r&0\\
    u_{z,r}&0&1+u_{z,z}\end{array}\right),
\end{align}
with $u_{i,j}\equiv\partial_j u_i$. In order to estimate roughly the order of the nonlinear coefficient, we avoid the derivatives and keep only $u_r/r$. Furthermore, we avoid $u_z$, which is much smaller than $u_r$ in the conventional RBM. Thus, the equation of motion is given by
\begin{align}
\rho \ddot{u}_r=&\nabla \sigma=\nabla \left(\frac{1}{|F|}FSF^t\right),\\
S_{ij}=&\lambda(\mathrm{Tr[E]})\delta_{ij}+2\mu E_{ij},
\end{align}
where $\lambda$ and $\mu$ are Lam\'e constants.
It can be approximated to
\begin{align}
    \rho\ddot{u}_r=-E_0\frac{u_r}{r^2}-2E_0\frac{u_r^3}{r^4}+\mathcal{O}(u_r^2,u_r^4),
\end{align}
where $E_0=\lambda+2\mu$ is the Young modulus, and $\mathcal{O}(u_r^2,u_r^4)$ denotes the second- and fourth-order terms that do not affect the rotating dynamics in the mechanical mode. This equation can be transformed to the mode equation via decomposition $u_r(r,z,t)=\phi_r(r,z)U(t)$ with $\mathrm{max}_{r,z}|\phi_{r}(r,z)|=1$. Finally, we obtain
\begin{align}
    \ddot{U}=-\Omega^2U+\alpha_G U^3
\end{align}
with 
\begin{align}
    \alpha=&\frac{2E}{m_\mathrm{eff}}\int  \frac{\phi_r^4}{r^4} dV,\\
    m_\mathrm{eff}=&\rho_0\int \phi^2_r(r)dV \nonumber.\\
\end{align}
By using the practical parameters in our device, $R_0=40$ $\mathrm{\mu m}$, $\beta=625$ $\mathrm{m^-1}$, $\rho_0=2648$ $\mathrm{kg/m^3}$, $E_0=72$ GPa, we estimate $\alpha_G=6.3\times10^{25}$ $\mathrm{m^{-2}s^{-2}}$, which is eight orders of magnitude smaller than $\alpha_\mathrm{OM}$. Such a small geometrical nonlinearity is consistent because of the small deformation in the RBMs.

\newpage

\section{Fitting Equation for Ring-Down with Nonlinear Damping}
To derive the ring-down dynamics in the presence of nonlinear damping, we begin with the forced nonlinear oscillator

\begin{equation}
    \ddot{x} + \left(\Gamma_{\mathrm{m}} + \beta x^{2}\right)\dot{x} + \Omega_{\mathrm{m}}^{2} x 
    = F \cos(\omega t).
\end{equation}
After the external drive is switched off (\(F=0\)), the equation of motion reduces to
\begin{equation}
    \ddot{x} + \left(\Gamma_{\mathrm{m}} + \beta x^{2}\right)\dot{x} + \Omega_{\mathrm{m}}^{2} x = 0.
    \label{eq:NonlinearDampEOM}
\end{equation}
This describes a free harmonic oscillator with intrinsic linear damping \(\Gamma_{\mathrm{m}}\) and amplitude-dependent nonlinear damping \(\beta x^{2}\). Assuming weak nonlinearity (\(\beta\) small), the motion can be represented by slowly varying envelope form
\begin{equation}
    x(t) = B(t)e^{-i\Omega_{\mathrm{m}} t} + \mathrm{c.c.},
\end{equation}
where envelope \(B(t)\) satisfies
\(\dot{B}\ll \Omega_{\mathrm{m}} B\) and \(\ddot{B}\ll \Omega_{\mathrm{m}}\dot{B}\).
Under this approximation, $\dot{x}(t)$ and $\ddot{x}(t)$ are, respectively, given by
\begin{eqnarray}
    \begin{gathered}
    \dot{x}(t)=\left(\dot{B}-i \Omega_{\mathrm{m}} B\right) e^{-i \Omega_{\mathrm{m}} t} + \mathrm{c.c.}, \\
    \ddot{x}(t)=\left(\ddot{B} -2 i \Omega_{\mathrm{m}} \dot{B}-\Omega_{\mathrm{m}}^2 B\right) e^{-i \Omega_{\mathrm{m}} t} + \mathrm{c.c.}
    \end{gathered}
\end{eqnarray}
For the nonlinear damping term, we have
\begin{eqnarray}
    x^2 \dot{x}
    &=& \left(B^2 e^{-2 i \Omega_{\mathrm{m}} t}+2\left|B\right|^2 + B^{*2} e^{2 i \Omega_{\mathrm{m}} t}\right) \left[ \left(\dot{B}-i \Omega_{\mathrm{m}} B\right) e^{-i \Omega_{\mathrm{m}} t} + \mathrm{c.c.} \right].
\end{eqnarray}
Extracting the resonance (positive-frequency) component yields
\begin{eqnarray}
    \left.x^2 \dot{x}\right|_{+} &=& 2\left|B\right|^2 \left(\dot{B}-i \Omega_{\mathrm{m}} B\right) e^{-i \Omega_{\mathrm{m}} t}
    + B^2 e^{-2 i \Omega_{\mathrm{m}} t} \left(\dot{B}^*+i \Omega_{\mathrm{m}} B^*\right) e^{i \Omega_{\mathrm{m}} t},
\end{eqnarray}
then the dominant slowly varying component becomes
\begin{eqnarray}
    \left.x^2 \dot{x}\right|_{+}
    &\approx& -i\Omega_{\mathrm{m}}\left|B\right|^2 B e^{-i \Omega_{\mathrm{m}} t}.
\end{eqnarray}
The nonlinear envelope equation can be rewritten as
\begin{eqnarray}
  \dot{B}
    &=&- \frac{\Gamma_{\mathrm{m}}}{2} - \frac{\beta}{2}\left|B\right|^2 B.
\end{eqnarray}
Next, we rewrite the real displacement in terms of a complex envelope as
\begin{equation}
    x(t)=B(t)e^{-i\Omega_{\mathrm m} t}+\mathrm{c.c.},
    \label{eq:x_complex_envelope}
\end{equation}
which guarantees that $x(t)$ is real.  On the other hand, the same motion can be written in real-amplitude form as
\begin{equation}
    x(t)=b(t)\cos(\Omega_{\mathrm m}t+\phi(t)).
    \label{eq:x_real_amplitude}
\end{equation}
Using $\cos\theta=(e^{i\theta}+e^{-i\theta})/2$ and comparing Eqs.~(\ref{eq:x_complex_envelope}) and
(\ref{eq:x_real_amplitude}), we identify
\begin{equation}
    B(t)=\frac{b(t)}{2}e^{-i\phi(t)}.
    \label{eq:B_in_terms_of_b}
\end{equation}
Therefore, real oscillation amplitude $b(t)$ and the modulus of the complex envelope, $B(t)$, are related by
\begin{equation}
    b(t)=2|B(t)|.
    \label{eq:b_2absB}
\end{equation}
In the following, we use Eq.~(\ref{eq:b_2absB}) to convert the envelope equation derived for $B(t)$ into the
fitting function for the experimentally measured real amplitude, $b(t)$. We obtain the envelope equation
\begin{equation}
    \dot{b}(t) = -\frac{1}{2}\Gamma_{\mathrm{m}}\, b(t) - \frac{1}{8}\beta\, b^{3}(t).
    \label{eq:bDot}
\end{equation}
Let \(\Gamma_{\mathrm{m}}' = \Gamma_{\mathrm{m}}/2\) and \(\beta' = \beta/8\).  
Then Eq.~(\ref{eq:bDot}) becomes
\begin{equation}
    \frac{db(t)}{dt} = -\Gamma_{\mathrm{m}}' b(t) - \beta' b^{3}(t).
\end{equation}
Separating variables yields
\begin{equation}
    \frac{db}{(\Gamma_{\mathrm{m}}' + \beta' b^{2})b} = - dt.
    \label{eq:Separation}
\end{equation}
Using partial-fraction decomposition,
\begin{equation}
    \int \frac{db}{(\Gamma_{\mathrm{m}}' + \beta' b^{2})b}
    = \frac{1}{\Gamma_{\mathrm{m}}'}\ln b 
      - \frac{1}{2\Gamma_{\mathrm{m}}'}\ln(\Gamma_{\mathrm{m}}' + \beta' b^{2}) + C_1.
\end{equation}
Integrating Eq.~(\ref{eq:Separation}) gives
\begin{equation}
    \frac{1}{2\Gamma_{\mathrm{m}}'}
    \ln\!\left( \frac{b^{2}}{\Gamma_{\mathrm{m}}' + \beta' b^{2}} \right)
    = - t + C.
\end{equation}
Exponentiating,
\begin{equation}
    \frac{b^{2}}{\Gamma_{\mathrm{m}}' + \beta' b^{2}}
    = C'' e^{-2\Gamma_{\mathrm{m}}' t}.
\end{equation}
Applying initial condition \(b(0)=b_{0}\),
\begin{equation}
    b^{2}(t)
    = \frac{b_{0}^{2} e^{-2\Gamma_{\mathrm{m}}' t}}
           {1 + (\beta' b_{0}^{2}/\Gamma_{\mathrm{m}}')\left( 1 - e^{-2\Gamma_{\mathrm{m}}' t} \right)}.
\end{equation}
Thus the amplitude envelope becomes
\begin{equation}
    b(t)
    = \frac{
        b_{0} e^{-\Gamma_{\mathrm{m}}' t}
      }{
        \sqrt{
            1 + (\beta' b_{0}^{2}/\Gamma_{\mathrm{m}}')
            \left( 1 - e^{-2\Gamma_{\mathrm{m}}' t} \right)
        }
      }.
\end{equation}
Re-substituting \(\Gamma_{\mathrm{m}}'=\Gamma_{\mathrm{m}}/2\) and \(\beta'=\beta/8\), we obtain the final fitting formula used in the analysis:
\begin{equation}
b(t)=
\frac{b_0 e^{-\frac{\Gamma_{\mathrm{m}}}{2} t}}
{\sqrt{1+R_{\mathrm{nl}}
\left(1-e^{-\Gamma_{\mathrm{m}} t}\right)}},
\qquad
R_{\mathrm{nl}}
=
\frac{\beta b_0^2}{4\Gamma_{\mathrm{m}}}.
\label{eq:FinalRingdown}
\end{equation}
Equation~(\ref{eq:FinalRingdown}) captures both the linear exponential decay at small amplitude and the enhanced or reduced decay due to nonlinear damping at large amplitude, and is used to fit the mechanical ring-down data.

\section{Linear optomechanical response}
\begin{figure}[h]
\centering
\includegraphics[width=0.6\linewidth]{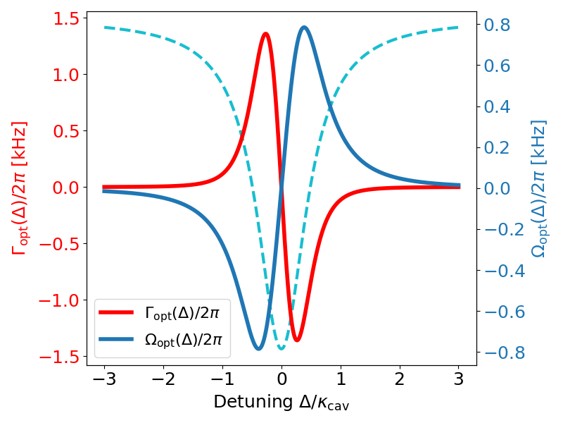}
\caption{
Linear (first-order) optomechanical damping (red) and the optical spring effect (blue) vs. laser detuning $\Delta$. The light-blue dashed line indicates the optical cavity transmission.
\label{fig:S4}}
\end{figure}

Using the parameters given in Sec.~\ref{subsec:1-4}, we calculate the laser detuning dependence of the linear (first-order) optomechanical effects, namely the optomechanical damping and the mechanical resonance frequency shift (optical spring effect). The results are shown in Fig.~\ref{fig:S4}. These results are consistent with the standard theory of cavity optomechanics, as reviewed in Ref.~\citep{aspelmeyer2014cavity}.
We note that the sign of the nonlinear damping coefficient shown in Fig.~1(b) of the main text is opposite to that of the linear optomechanical damping. This sign reversal originates from the phase structure of the optomechanical response. In the case of linear optomechanical damping, the delayed optical response introduces a phase shift of approximately $\pi/2$ relative to the mechanical motion, resulting in a dissipative force proportional to the velocity. In contrast, the nonlinear damping arises from a cubic optomechanical response, effectively involving three such phase contributions. This leads to an accumulated phase shift of approximately $3\pi/2$, thereby reversing the sign of the dissipative contribution compared to the linear case.

\section{Photothermally induced nonlinear damping}
Based on our additional measurements shown in Figs. 3(b)–3(d) of the main text, we observe that the sign of the nonlinear damping reverses depending on the laser detuning. This behavior provides strong experimental evidence that the observed nonlinear damping originates from optomechanical coupling, since photothermal effects do not exhibit such a sign reversal between the red- and blue-detuned regimes. In addition, the good agreement between the experimentally observed nonlinear damping coefficients and the theoretical estimates based on optomechanical coupling further supports the optomechanical origin of the observed nonlinearity.

For completeness, we have also carefully examined the possibility that photothermal effects contribute to the observed nonlinear damping under our experimental conditions. In our system, two primary mechanisms can give rise to photothermal-induced nonlinear damping \citep{atalaya2016nonlinear}:
(1) thermoelastic nonlinear damping, and  
(2) Akhiezer-type nonlinear damping.

(1) Thermoelastic nonlinear damping originates from the relaxation of spatially non-uniform temperature fields via thermal diffusion. The thermal response time $\tau_{\mathrm{th}}$ of our device is on the order of several hundred $\mu$s, satisfying the condition $\Omega_{\mathrm{m}}\tau_{\mathrm{th}} \gg 1$. In this regime, the temperature field cannot follow the mechanical oscillation and the modulation of the pump laser intensity used to drive the mechanical motion. As a result, the photothermal effect is effectively quasi-static with respect to the mechanical motion, and the dynamic photothermal backaction is strongly suppressed. Consequently, nonlinear damping arising from the thermoelastic mechanism is not expected to be significant in our experiments.

(2) Akhiezer-type nonlinear damping arises from the nonequilibrium relaxation of thermal phonon distributions modulated by mechanical vibrations. The relevant timescale is the relaxation time of thermal phonons, $\tau_{\mathrm{phonon}}$. In silica, this relaxation time is reported to be on the order of $\tau_{\mathrm{phonon}} \sim 1$ ps \citep{vacher2005anharmonic}. Under our experimental conditions, this gives $\Omega_{\mathrm{m}}\tau_{\mathrm{phonon}} \sim 2\pi\times4.8\times 10^{-5}\ll 1$. Since the magnitude of Akhiezer-type nonlinear damping scales as $\Omega_{\mathrm{m}}\tau_{\mathrm{phonon}} / (1 + \Omega_{\mathrm{m}}^2\tau_{\mathrm{phonon}}^2)$, the system operates far from the condition $\Omega_{\mathrm{m}}\tau_{\mathrm{phonon}} \sim 1$ where this mechanism is maximized. Therefore, the contribution of this mechanism is also strongly suppressed.

From these considerations, we conclude that photothermal effects do not provide a dominant contribution to the observed nonlinear damping under our experimental conditions. Instead, the observed nonlinear damping is primarily attributed to optomechanical coupling.


\end{document}